\documentclass[aps,prb,twocolumn,superscriptaddress]{revtex4-2}
\usepackage{graphicx}
\usepackage{amsmath}
\usepackage{amssymb}
\usepackage{mathtools}
\usepackage{xcolor}
\usepackage{braket}
\usepackage[colorlinks=true, citecolor=blue, urlcolor=blue, linkcolor=blue,bookmarks=false,hypertexnames=true]{hyperref} 
\usepackage{appendix}

\usepackage{array}
\usepackage[column=O]{cellspace}
\newcolumntype{P}[1]{>{\centering\arraybackslash}p{#1}}

\begin{document}
\graphicspath{}
\preprint{APS/123-QED}

\title{Quantum simulator of extended bipartite Hubbard model with broken sublattice symmetry: magnetism, correlations, and phase transitions}

\author{Yasser Saleem}
\affiliation{%
 Department of Physics, University of Ottawa, Ottawa, Canada  
}
\author{Amintor Dusko}%
\affiliation{%
 Department of Physics, University of Ottawa, Ottawa, Canada 
}

\author{Moritz Cygorek}
\affiliation{%
  Department of Physics, University of Ottawa, Ottawa, Canada
}

\author{Marek Korkusinski}
\affiliation{
National Research Council of Canada, Ottawa, Canada}
\affiliation{%
  Department of Physics, University of Ottawa, Ottawa, Canada
}
%
\author{Pawel Hawrylak}
\affiliation{%
 Department of Physics, University of Ottawa, Ottawa, Canada
}%

\date{\today}

\begin{abstract}
We describe here a quantum simulator of extended bipartite Hubbard model with broken sublattice symmetry. The simulator consists of a structured lateral gate confining two dimensional electrons in a quantum well into artificial minima arranged in a hexagonal lattice. The sublattice symmetry breaking is generated by forming an artificial triangular graphene quantum dot (ATGQD) with zigzag edges. The resulting extended Hubbard model generates tunable ratio of tunneling strength  to electron-electron interactions andbof sublattice symmetry with control over shape. The validity of the simulator is confirmed for small systems using mean-field and exact diagonalization many-body approaches which show that the ground state changes from a metallic to an antiferromagnetic (AF) phase by varying the distance between sites or depth of the confining potential. The one-electron spectrum of these triangular dots contains a macroscopically degenerate shell at the Fermi level. The shell persists at the mean-field level for weak interactions (metallic phase)  but disappears for strong interactions, in the AF phase. We determine the effects of electron-electron interactions on the ground state, the total spin, and the excitation spectrum as a function of filling of the ATGQD. We find that the half-filled charge neutral shell leads to a partially spin polarized state in both metallic and AF regimes in accordance with Lieb’s theorem. In both regimes a relatively large gap separates the spin polarized ground state to the first excited many-body state at half filling of the degenerate shell. By adding or removing an electron, this gap drops dramatically, and alternate total spin states emerge with energies nearly degenerate to a spin polarized ground state.
\end{abstract}

\maketitle
\section{Introduction}
There is currently interest in understanding electronic properties of strongly correlated quantum materials often modelled by an extended Hubbard model. It is expected that progress in solutions to this intractable problem  may be achieved with quantum simulators. Here we describe  a proposal of a quantum simulator of extended bipartite Hubbard model with broken sublattice symmetry inspired by graphene. Much progress in quantum simulators has been achieved with cold atoms and trapped ions \cite{lloyd1996universal,bloch2005ultracold,
jaksch2005cold,ortner2009quantum,mazurenko2017cold,kuhr2016quantum,weimer2010rydberg,barreiro2011open,hempel2018quantum,islam2011onset,leykam2018artificial}. Progress in solid state and photonic based simulators \cite{salfi2016quantum,buluta2009quantum,aspuru2012photonic,bernien2017probing,cai2013large,dusko2018adequacy,singha2011two,hofling2018,park2009making,singha2011two,li2021anisotropic,shi2019gate,forsythe2018band,gibertini2009engineering,kylanpaa2016stability,uehlinger2013artificial,rasanen2012electron,bloch2014} is enabled by progress in new materials, including quasi-two-dimensional  electronic systems (2DES) in semiconductor heterojunctions \cite{ep2ds1999,ep2ds2009,piquero2017precise,wang2014manipulation} and graphene. 
The isolation of a single carbon layer, graphene, introduced a new 2DES with unusual electronic properties, including the zero energy band gap, relativistic nature of quasiparticles, sublattice pseudospin and two non-equivalent valleys\cite{wallace1947band,novoselov2004electric,zhang2005experimental,zhou2006first,neto2009electronic,novoselov2005two,guttinger2010spin,wang2014manipulation}. Finite lateral size quantization of graphene opens up an energy gap making graphene a 2D atomically thin semiconductor with a gap tunable from THz to UV \cite{saleem2019oscillations,Guclu_graphene,gucclu2011electric,gucclu2009magnetism,potasz2010zero,potasz2012electronic,ponomarenko2008chaotic,lu2011transforming,ezawa2010dirac,wunsch2008electron}.

Some of these known properties of graphene and their previous works inspire our approach to create a simulator based on  artificially structured gates on top of a 2D electron gas in a field effect transistor. Artificial graphene structures have been realized already using photonic lattices, nano-patterning, modulation doping, and scanning probe methods for atomic manipulation on metal surfaces.\cite{bloch2005ultracold,gibertini2009engineering,forsythe2018band,park2009making,singha2011two,li2021anisotropic,peleg2007conical,gomes2012designer,uehlinger2013artificial,shi2019gate,rasanen2012electron,kylanpaa2016stability,drost2017topological,slot2017experimental}. Here we propose a quantum simulator of graphene inspired bipartite extended Hubbard model with broken sublattice symmetry. The working of the simulator is confirmed for small systems using mean-field and exact diagonalization methods which show how magnetism, correlations, and phase transitions emerge, as parameters of the simulator are varied.

We focus here on a very important property of hexagonal lattice, i.e., the presence of two triangular sublattices. The sublattice symmetry can be broken in finite, triangular quantum dots with zigzag edges, and a macroscopic band of degenerate one electron states emerges at the Fermi level. Lieb predicted that a bipartite Hubbard model with broken sublattice symmetry will have a finite magnetic moment\cite{lieb1989two}. Indeed, for a half-filled system, Ezawa, Fernandes Rossier, Kaxiras, Potasz, and others  \cite{ezawa2007metallic,fernandez2007magnetism,wang2009topological,potasz2012electronic,gucclu2011electric,potasz2010zero, gucclu2009magnetism} found the ground state to be partially spin polarized in agreement with Lieb’s theorem \cite{lieb1989two} and with exchange interaction being responsible for aligning spins of electrons on the zero-energy degenerate shell. This additional polarization was found to be proportional to the imbalance of the two sublattices and to degeneracy of the shell, with extra electron spins localized largely at the edge of the triangular structure. Such small triangular structures were realized experimentally and confirmed theoretical predictions at half-filling \cite{pavlivcek2017synthesis,mishra2020topological,su2019atomically}. The presence of a degenerate macroscopic shell at the Fermi level, analogous to the lowest Landau level , was shown to lead to strong correlations in the ground state. It was found that the addition of a single additional electron to a half-filled zero-energy shell destroyed the spin polarization\cite{gucclu2009magnetism,potasz2012electronic}.

There are several advantages in transferring the physics of graphene to an artificial graphene-like quantum simulator where carbon atoms are replaced by gated quantum dots and the structure is embedded in a semiconductor host. These advantages include tunable distance between the dots and depth of confining potential, programmable lattice symmetry and termination (i.e. edge type), tunable electron-electron interactions and interdot tunneling \cite{kylanpaa2016stability,rasanen2012electron,singha2011two}. Further advantages of artificial graphene include the ability to control values of $U/t$ in a bipartite Hubbard model, which is not possible with graphene \cite{wehling2011,wehling2014dirac}. Such control would allow to demonstrate different electronic phases, including transition from semi-metal to an AF insulator \cite{sorella1992semi,salfi2016quantum} . Additionally, triangular graphene quantum dots are susceptible to edge reconstruction as studied by Voznyy et. al. \cite{voznyy2011effect}. Edge reconstruction is responsible for smearing out the distinction between sublattices, and reduces the quantum dot symmetry. These combined features can destroy the magnetic properties of the system. The difficulty of edge reconstruction  is overcome in an artificial system, where the edge is determined by the external gate. Another important advantage of artificial graphene is that, unlike in graphene, in AG a single electron can be placed in the system in order to probe the single particle spectrum, directly demonstrating the existence of a zero energy shell and relating  it to many-electron properties.

In this paper we report on the study of the quantum simulator of an ATGQD with zig-zag edges. The quantum dot is formed by a structured metallic gate generating a lattice of potential minima arranged to form a hexagonal lattice forming a triangle with zigzag edges in a semiconductor quantum well with 2D electron gas. We compute tunneling matrix elements and Coulomb matrix elements as a function of AG parameters. We confirm the existence of a macroscopic shell of degenerate states as found in triangular graphene quantum dots. Then, we compute the Hartree-Fock (HF) spectrum for a half-filled system as a function of the strength of Coulomb interactions as measured by the ratio of Coulomb on-site repulsion $U$ to nearest neighbor tunneling matrix element $t$. We find a metallic phase with a zero-energy shell at the Fermi level for weak Coulomb interactions and an insulating AF phase, without a shell, for strong Coulomb interactions. We find the ground state of a half-filled system to be partially spin polarized, in agreement with Lieb’s Theorem. With explicit calculations of $U/t$ we determine how to transition between these two phases. We use exact diagonalization techniques to show the spin depolarization as a function of  filling factors for metallic and AF phases. These results are presented for a small size ATGQD where numerical calculations are possible and test the viability of the Quantum simulator for large, intractable, systems. 

The paper is organized as follows: in Section II, we describe the model, geometry and Hamiltonian, in section III we describe the one electron tight-binding model. In section IV we discuss the many-body Hamiltonian and the calculation of Coulomb matrix elements. Section V discusses the HF ground state for weak,metallic,  and strong, AF,  Coulomb interactions. In Section VI we add correlations within a TB-HF-configuration-interaction method, and demonstrate that the inclusion of an extra electron collapses the energy gap and depolarizes the electronic system.
\begin{figure}[tb]
    \centering
    \includegraphics[scale=0.75]{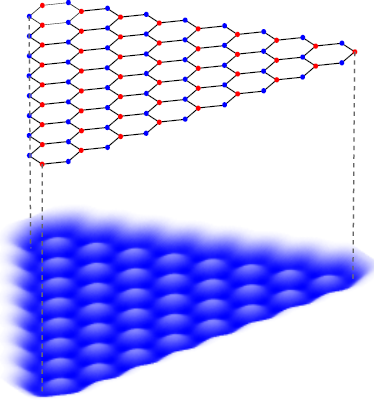}
    \caption{Potential profile of ATGQD with zigzag edges, with {$N=97$} sites separated by  {$a=15$ nm}. There is broken sublattice symmetry in this system in which the number of A sites does not equal the number of B sites, as seen in the difference in red and blue dots.}
    \label{fig:fig1}
\end{figure}
\section{MODEL OF TRIANGULAR ARTIFICIAL GRAPHENE QUANTUM DOT}
We start with electrons confined to a quantum well described by a potential {$V(z)$} where z is the growth direction. A metallic gate is deposited on a surface at a distance “$D$” from the quantum well. The potential on the gate generates a potential {$V(\vec{r}-\vec{R})$} in the plane of a quantum well laterally confining an electron at {$\vec{r}$} in the vicinity of position {$\vec{R}$}. An artificial graphene structure is defined by structuring the metallic gate resulting in an array of $N$ confining potentials, artificial atoms, positioned on a hexagonal lattice of potential minima at {$\vec{R_i}$} separated by a distance $"a"$. We next introduce a back gate from which electrons are drawn into the artificial graphene structure leaving behind a positive, compensating charge described by a gate potential {$V_g$}. Hence, the artificial graphene Hamiltonian describing {$N_e$} electrons in an array of $N$ potential traps in a quantum well, in the presence of a compensating gate potential and including electron-electron interactions is given by
\begin{multline}
    \scalebox{1.0}{$
      H = \sum\limits_{i=1}\limits^{N_e}\left[-\frac{\hbar^2}{2m^*}\nabla^2_i+\sum\limits_{j=1}\limits^{N}v(\vec{r_i}-\vec{R_j})+v(z_i) \right.$}
     \\
    \scalebox{1.0}{$ \left. +V_g\left(\vec{r_i},z_i\right) \right]+\sum\limits_{i<j}\limits^{N_e}\frac{e^2}{\kappa|\vec{r_i}-\vec{r_j}|},$}
        \label{eq:Hamiltonian1}
    \end{multline}
Here, we sum over {$N_e$} electrons with effective mass {$m^*$} in the field produced by  an array of $N$ sites, interacting with a gate, confined to a quantum well by potential {$v(z_i)$} and interacting with each other via Coulomb potential screened by a background dielectric function {$\kappa$}. 
The potential minima {$v(\vec{r_i}-\vec{R_j})$} are finite and smooth confining potentials. Here we describe these potentials by a Gaussian potential  
\begin{equation}
    v(\vec{r_i}-\vec{R_j})=-V_0e^{-\frac{|\vec{r_i}-\vec{R_j}|^2}{d^2}},
    \label{eq:GaussianPot}
\end{equation}
with depth {$V_0$}, confinement length $d$, localized in the plane of a quantum well at {$\vec{R_j}$}. The one electron potential {$V(r)=\sum_jv(\vec{r}-\vec{R_j})$} of the artificial graphene structure studied here is shown in Fig. \ref{fig:fig1}. Different structures with different size and shape can be constructed analogously. Here, there are {$N=97$} sites, with each modelled as a Gaussian confining potential, with a depth {$V_0=300$ meV}, confinement length {$d=10$  nm} and separation of {$a=15$ nm}. We see that the confining potential forms a triangular quantum well, with minima arranged on a hexagonal lattice, with visible benzene like rings and terminated  by zigzag edges. Such a structure is an example of a bipartite lattice with broken sublattice symmetry, and as such Lieb’s theorem \cite{lieb1989two} will apply and play a critical role in determining the nature of the ground state.
\section{ONE ELECTRON SPECTRUM IN THE TIGHT-BINDING MODEL}
We now introduce one electron into the artificial graphene structure shown in Fig. \ref{fig:fig1}. Following the model 
described in Section II, the single particle Hamiltonian is given by
\begin{equation}
   H_0=-\frac{\hbar^2}{2m^*}\nabla^2+\sum_jv(\vec{r}-\vec{R_j})+v(z),
    \label{eq:TBHam}
\end{equation}
where the sum over $j$ extends over $N$ sites, and {$v(\vec{r}-\vec{R_j})$} is given in Eq. (\ref{eq:GaussianPot}). Here $v(z)$ is a potential of an infinite quantum well with width $0.1a$. Throughout this article, we assume strong confinement in the $z$-direction, so that the eigenstates factorize into an in-plane part and part in the z-direction, which we assume to be the lowest state of an infinite quantum well {$\xi\left(z\right) = \sqrt{\frac{2}{L}}\sin{\frac{\pi z}{L}}$}. The lateral confining potential is smooth and parabolic at low energies. Hence we expand the in-plane part of the wavefunction {$\varphi$} for a state {$\nu$} in terms of two-dimensional harmonic oscillator eigenfunctions {$\alpha$} centered on atom $j$:
\begin{equation}
   \varphi^\nu=\sum_{j,\alpha}A^\nu_{j\alpha}\phi^0_{j\alpha}.
       \label{eq:NonOWF}
\end{equation}
Acting with {$H_0$} in Eq.(\ref{eq:TBHam}) on the wavefunction Eq.(\ref{eq:NonOWF}), and noting that we can decompose the Gaussian in terms of a parabolic confining potential plus a correction as {$v(\vec{r}-\vec{R_j})=v_j^{ho}+\delta V_j$}  with {$v_j^{ho}=-V_0+\frac{\overline{\omega}^2}{4}|\vec{r}-\vec{R_j}|^2$} and {$\overline{\omega}^2=\frac{4}{d^2}V_0$}, we get matrix elements of the Hamiltonian in Eq. (\ref{eq:TBHam}) given by 
\begin{multline}
    \scalebox{1.0}{$
    H_{m\beta,l\alpha}=\epsilon^0_\beta S_{m\beta,l\alpha}+\braket{\phi^0_{m\beta}|\delta V_m|\phi^0_{l\alpha}}$}
     \\
    \scalebox{1.0}{$+\sum_{j\neq m}\braket{\phi^0_{m\beta}|v^{ho}_j+\delta V_j|\phi^0_{l\alpha}}.$}
    \end{multline}
    \label{eq:TBMatrixElements}
Here { $S_{m\beta, l\alpha}=\langle \phi^0_{m\beta} | \phi^0_{l\alpha} \rangle$ } are overlap matrix elements for orbitals {$\beta,\alpha$}   localized on sites {$m, l$}. 
Since the harmonic oscillator states on different sites defined in Eq.(\ref{eq:NonOWF}) are not orthogonal, we orthogonalize the basis by solving the generalized eigenvalue problem given by 
\begin{equation}
  \textbf{\~{H}}\textbf{B}^\nu=\epsilon^\nu\textbf{B}^\nu,
    \label{eq:GeneralizeMatrixEqn}
\end{equation}
where {$\textbf{B}^\nu=\textbf{S}^{1/2}\textbf{A}^\nu$} and {$\textbf{S}$} is the overlap matrix. The corresponding renormalized Hamiltonian given by 
\begin{equation}
  \textbf{\~{H}}=\textbf{S}^{-\frac{1}{2}}\textbf{H}\textbf{S}^{-\frac{1}{2}}.
    \label{eq:GeneralizedHam}
\end{equation}
The wavefunction is now expanded in terms of orthogonal states {$\psi_{m\beta}$} localized on different sites
\begin{equation}
   \phi^\nu=\sum_{m,\beta}B^\nu_{m\beta}\psi_{m\beta}.
       \label{eq:intinerantWF}
\end{equation}
The orthogonal orbitals localized on site {$"m"$} are given explicitly by 
\begin{equation}
  \psi_{m\beta}=\sum_{l\alpha}(S^{-\frac{1}{2}})_{l\alpha,m\beta} \phi^0_{l\beta}.
       \label{eq:OrthogonalWF}
\end{equation}
\begin{figure}
    \centering
    \includegraphics[width=\linewidth]{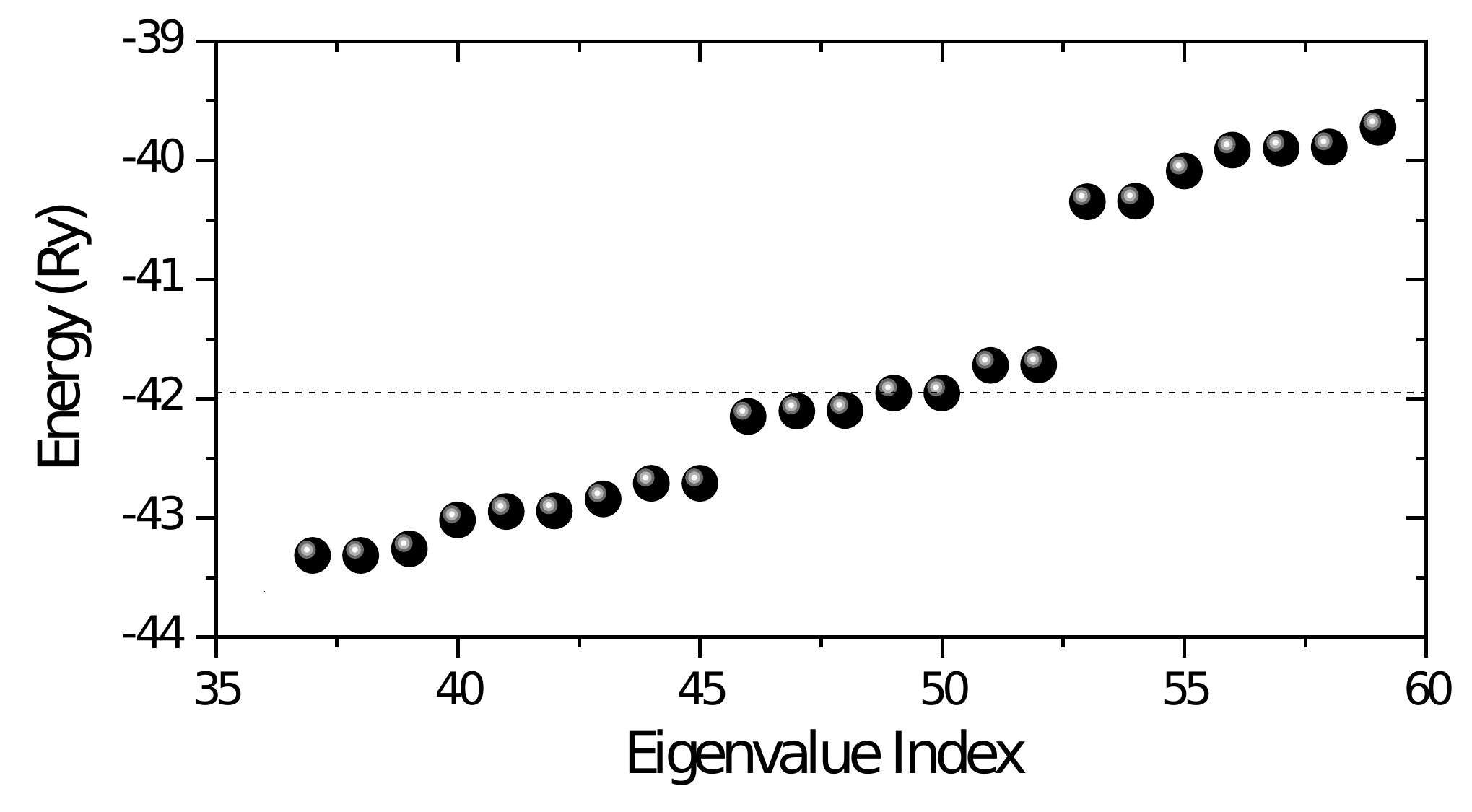}
    \caption{TB spectrum for AG Triangular quantum dot with zigzag edges near the Fermi level.}
    \label{fig:fig2}
\end{figure}
Fig. \ref{fig:fig2} shows the energy levels of a single electron obtained by diagonalizing Eq.(\ref{eq:GeneralizedHam}) for $N=97$ sites in the basis of 3 harmonic oscillator shells, $S$, $P$, and $D$ per site, with separation of potential minima corresponding to {$a=12.5$} nm, and all other parameters given in section II. We observe a well-defined shell of almost degenerate states at the Fermi level contained in the 1$S$-band. We note that the Fermi level corresponds to a half-filled electron system.  
A large gap separates the 1$S$ band of levels from the $P$, and $D$ derived levels and the single particle spectrum resembles a spectrum obtained by diagonalizing a tight binding model with nearest neighbor tunneling matrix elements \cite{gucclu2009magnetism}. Thus we limit ourselves to  only $S$ orbitals from here on in order to simplify many-body calculations. We take the hopping parameter “$t$” between nearest neighbour 1$S$ orbitals to be equal to half the bandwidth of the 1$S$ band. We find {$t=26.98$ meV} for site separation of {$a=12.5$ nm} and {$t=3.18$ meV} for site separation of {$a=15$ nm}. In Fig.\ref{fig:fig2} we see a shell of nearly degenerate zero-energy states at the Fermi level split by the introduction of next-nearest-neighbour hopping generated by the itinerant orbitals. Note that in artificial graphene, it is possible to have a single electron moving in the system of potential minima, and optical experiments, for example, could probe the existence of this zero energy shell.
\section{Many-body Hamiltonian}
With the orthogonalized orbitals {$\psi_{m\beta}$} limited to the 1$S$ band, we can write the many-body Hamiltonian, Eq. (\ref{eq:Hamiltonian1}) in the second quantized form as 
\begin{multline}
    \scalebox{1.0}{$
    H=\sum\limits_{i,\sigma}\epsilon_{i\sigma} c_{i\sigma}^\dag c_{i\sigma}+\sum\limits_{i,j,\sigma}t_{ij} c_{i\sigma}^\dag c_{j\sigma}+$}
     \\
    \scalebox{1.0}{$\frac{1}{2}\sum\limits_{i,j,k,l,\sigma,\sigma'}\braket{ij|V|kl}c_{i\sigma}^\dag c_{j\sigma'}^\dag c_{k\sigma'} c_{l\sigma} +\sum\limits_{i,\sigma}v_{ii}^g c_{i\sigma}^\dag c_{i\sigma},$}
     \label{eq:ManyBHam}
    \end{multline}
where  {$\epsilon_{i\sigma}$} are the onsite energies and {$t_{ij\sigma}$} are the hopping matrix elements computed above. {$v_{ii}^g$} corresponds to the back gate and {$\braket{ij|V|kl}$} are Coulomb matrix elements given by 
\begin{multline}
    \scalebox{1.0}{$
   \braket{ij|V|kl}= \int\int d\vec{r}_1d\vec{r}_2 \psi^*_i(\vec{r}_1) \xi_i^*(z_1) \psi^*_j(\vec{r}_2)\xi_j^*(z_2) $}
     \\
    \scalebox{1.0}{$\times\frac{2R_y}{|\vec{r}_1-\vec{r}_2+(z_1\hat{z}-z_2\hat{z})|}\psi_k(\vec{r}_2) \xi_k(z_2) \psi_l(\vec{r}_1)\xi_l(z_1),$}
       \label{eq:CoulombMatrixElements}
    \end{multline}
where the functions {$\psi_i$} are the 1$S$ localized and orthogonal orbitals which are defined in Eq.(\ref{eq:OrthogonalWF}). 
The functions {$\xi_i=\sqrt{\frac{2}{L}}sin(\frac{\pi z_i}{L})$} describe the lowest energy sub-band of an infinite quantum well confining the electrons in the $z$-direction. The back gate term is given by
\begin{equation}
  v_{ii}^g=\sum_j\frac{-\frac{2N_p}{N}}{\sqrt{(x_i-x_j)^2+(y_i-y_j)^2+d_{gate}^2}},
    \label{eq:PositiveBackground}
\end{equation}
where {$N_p$} is the number of positive charges on the gate. The model assumes that the number of positive charges on the gate {$N_p$} is equal to the number of electron charges in the system $N$ in order to enforce charge neutrality, and is uniformly smeared on the gate. 

Since the Hamiltonian in Eq.(\ref{eq:ManyBHam}) cannot be solved exactly, we start by solving the mean-field Hartree-Fock  problem first.

\section{Mean-field Hartree-Fock}
\begin{figure}
    \centering
    \includegraphics[width=\linewidth]{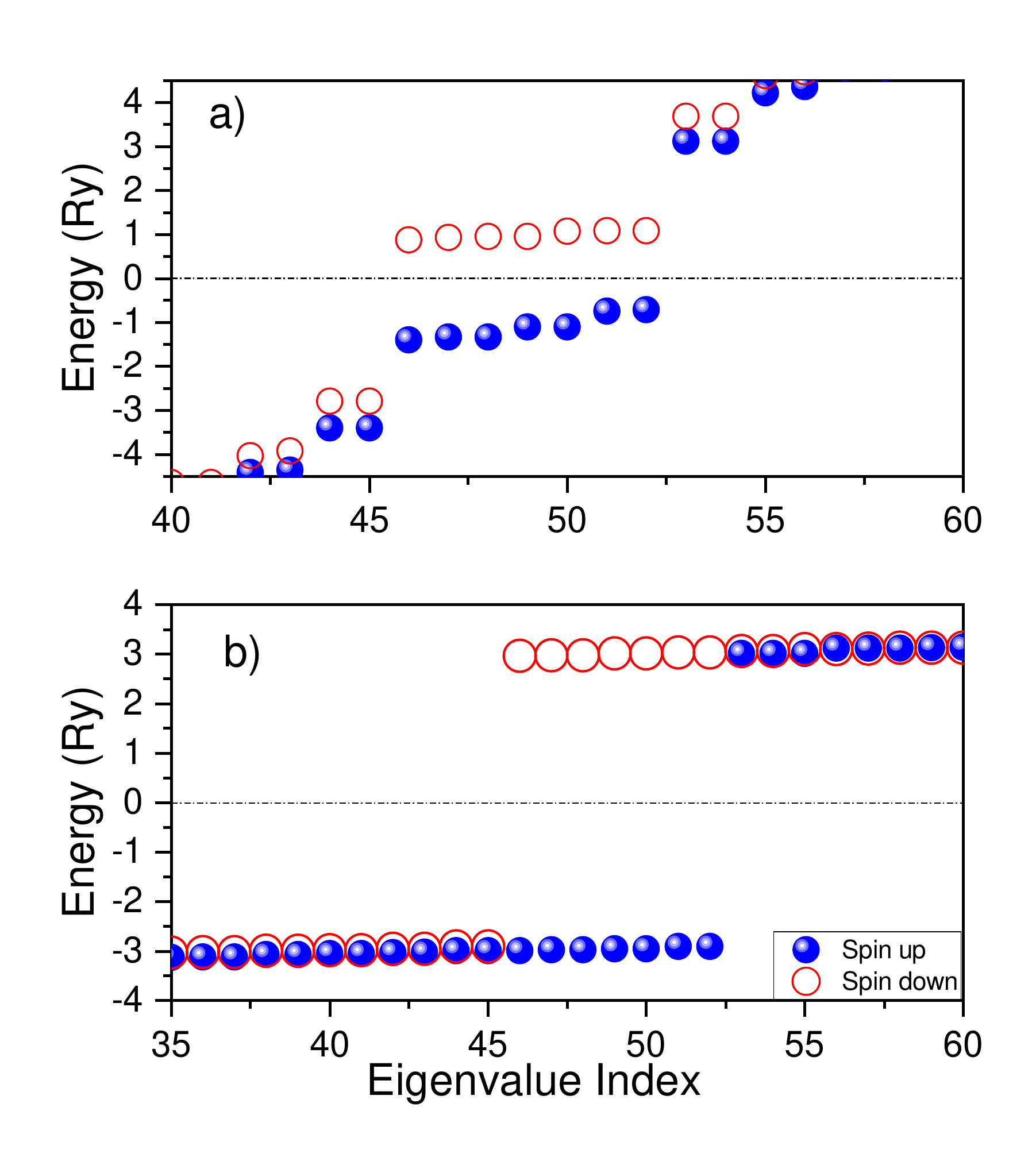}
    \caption{a) HF Spectrum for ground state of N=97 sites at half-filling with  lattice constant {$a=12.5$ nm}. We observe a shell of distinct degenerate states near the Fermi level, filling up of levels with spin up and down electrons, and a splitting of spin degeneracy of all levels  due to an imbalance of  up and down spins found in HF solution. All states below the Fermi level ${E_F=0}$ (middle line) are occupied, all states above are unoccupied. 
    b) HF spectrum for lattice separation {$a=15$ nm}. Note the disappearance of a degenerate shell at the Fermi level and emergence of a large gap proportional to Hubbard U separating the valence and conduction band states, AF spin ordering and  partially spin polarized ground state.}
    \label{fig:fig3}
\end{figure}
\begin{figure*}
    \centering
    \includegraphics[width=\linewidth]{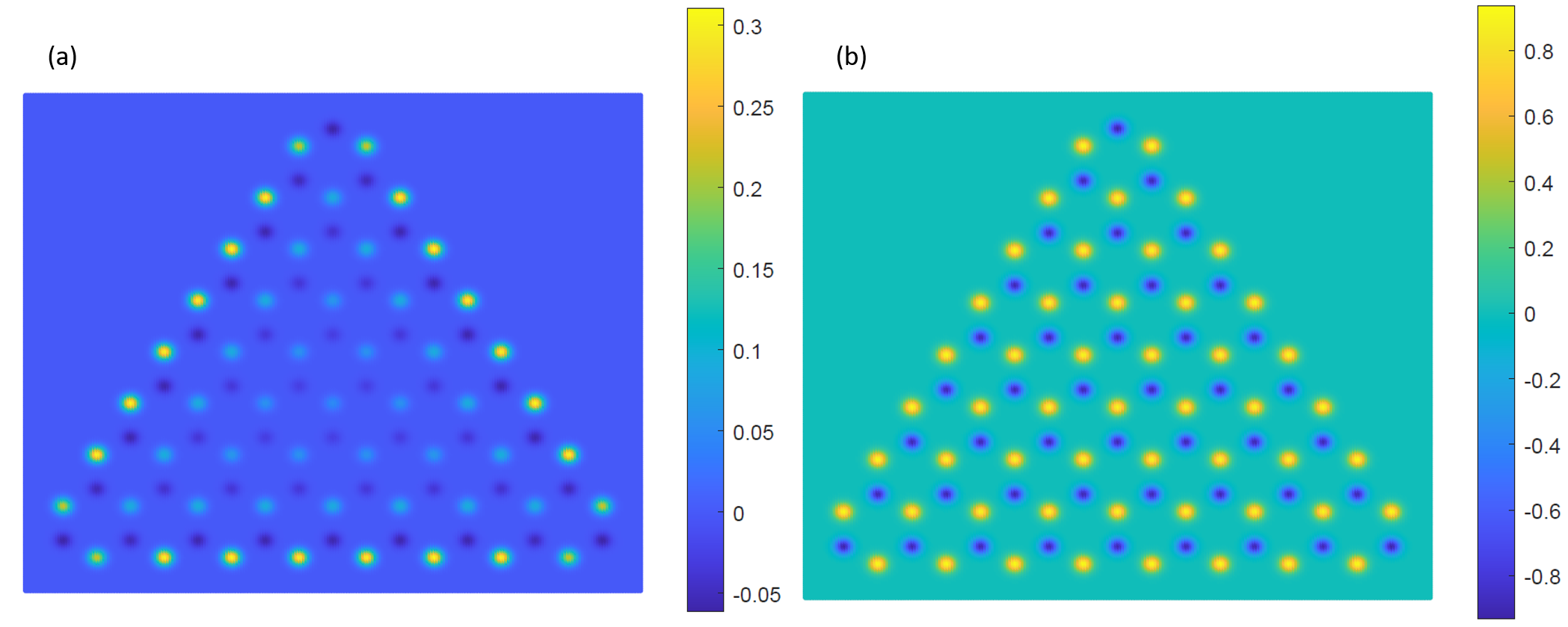}
    \caption{a) Spin density obtained for the spin polarized HF ground state for {$a=12.5$ nm}. We observe the system to be in the metallic phase, with the extra spins occupying the edge. b) Spin density for the spin polarized ground state for {$a=15$ nm} . We notice one spin is localized on one sublattice, while the other, is localized on the other sublattice. Due to a sublattice imbalance this as well leads to spin polarized electrons localized on the edge.}
    \label{fig:fig4}
\end{figure*}
The magnetic properties of artificial graphene, and the nature of the ground state for different electron numbers, can be studied by solving the Mean-field HF Hamiltonian obtained from Eq.(\ref{eq:ManyBHam}) and  given by:
\begin{multline}
    \scalebox{1.0}{$
    H^{ATGQD}_{MF}= \sum\limits_{i,l,\sigma}\tau^0_{il\sigma}c_{i\sigma}^\dag c_{l\sigma} $}
     \\
    \scalebox{1.0}{$ + \sum\limits_{i,j,k,l,\sigma,\sigma'} \left[\braket{ij|V|kl}-\braket{ij|V|lk}\delta_{\sigma \sigma'}\right]$}
     \\
    \scalebox{1.0}{$ \times \left(\rho_{jk\sigma'}-\rho^0_{jk\sigma'}\right) c_{i\sigma}^\dag c_{l\sigma} + \sum\limits_{i,\sigma}v_{ii}^gc_{i\sigma}^\dag c_{i\sigma}, $}
       \label{eq:MFH}
    \end{multline}
where {$\rho_{jk\sigma'}$} is the density matrix elements for the ATGQD, and {$\rho^0_{jk\sigma'}$} is the density matrix for the bulk system. The choice of {$\rho^0_{jk\sigma'}$} is discussed in appendix A, and its purpose is to enhance convergence. {$\tau^0_{il\sigma}$} is a tunneling matrix element which describes the properties of bulk artificial graphene in terms of the tunneling matrix element {$t_{il}$}  and  bulk density matrix {$\rho^0_{jk\sigma}$}. It is given by
\begin{equation}
  \tau^0_{il\sigma} = -t_{il\sigma}+\sum_{jk\sigma'}\left[\braket{ij|V|kl}-\braket{ij|V|lk}\delta_{\sigma \sigma'}\right]\rho^0_{jk\sigma'},
       \label{eq:tauil}
\end{equation}
Here the superscript $0$ on {$\tau^0_{il\sigma}$} is to denote that it is computed with the bulk density matrix elements. In addition to the onsite interaction terms, all direct terms are taken into account, as well as all exchange terms up to next nearest neighbours. The bulk density matrix elements are computed in the appendix for the AFM regime. They have been obtained by Potasz et al. \cite{potasz2012electronic} for the metallic regime. We note the Hamiltonian in Eq.\ref{eq:MFH} is symmetric with respect to spin, and thus can be diagonalized in separate subspaces for spin up and spin down, but with the spin up Hamiltonian having a dependence on the density of spin down electrons and vice versa. We focus on {$S_z\geq0$} and proceed to diagonalize Eq.(\ref{eq:MFH}), with results shown in Fig.\ref{fig:fig3} and Fig.\ref{fig:fig4}. 
Fig.\ref{fig:fig3}(a) shows the energy spectrum for spin up and down electrons in the metallic regime , $a=12.5$ nm. We see a spin splitting of levels due to a spin imbalance obtained in HF and consistent with Lieb's theorem. We observe a nearly degenerate shell at the Fermi level, with the blue spin up electrons  fully occupying a degenerate shell, leaving the red spin down levels completely empty above  the Fermi level. These extra spins are found to align on the edge of the triangle as seen in  Fig.\ref{fig:fig4}(a), with a uniform zero spin density away from the edges indicating a semi-metallic regime. 
We now proceed to strongly interacting regime by increasing the distance between lattice sites to $a=15$ nm. We find the ground state to be again partially spin polarized due to broken sublattice symmetry in agreement with Lieb’s theorem \cite{lieb1989two}. We however lose a distinct degenerate shell at the Fermi level, instead we find a large gap, proportional to Hubbard U, separating the valence and conduction band, suggesting an insulating phase  (Fig.\ref{fig:fig3}(b)). We observe an antiferromagnetic spin ordering in the bulk as shown in Fig.\ref{fig:fig4}(b) as expected in the large U/t regime \cite{sorella1992semi} and ferromagnetic ordering at the edges of the ATGQD.
\begin{figure*}
    \centering
    \includegraphics[width=1\linewidth]{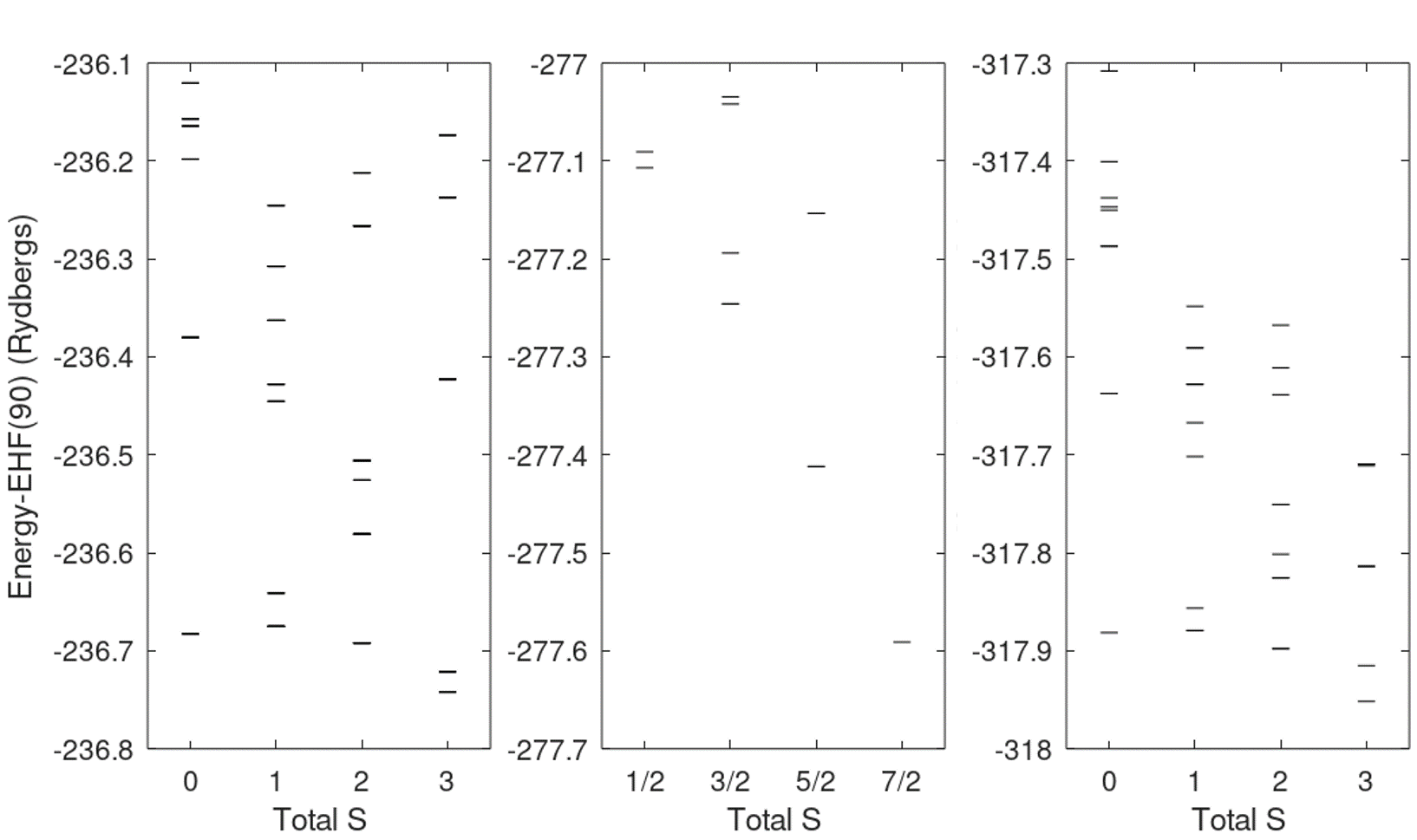}
    \caption{The low-energy spectra in the metallic phase for a) 97-1 electrons b) 97 electrons c) 97+1 electrons. For 97 electrons the ground state is partially spin polarized, and we see a large gap that separates this state from other spin states, but the introduction or removal of an electron collapses this gap and many spins states exist at very close energies to the spin polarized ground state for the {$N=97-1$} and {$N=97+1$} electron cases.}
    \label{fig:fig5}
\end{figure*}
\section{ELECTRONIC CORRELATIONS VIA CONFIGURATION INTERACTION}
\subsection{ELECTRONIC CORRELATIONS IN THE METALLIC PHASE}
We now turn to include electronic correlations. Let us begin with the semi-metallic phase. In the example of ATGQD with $N=97$ sites, the breaking of sublattice symmetry results in $N_d=7$ fold degenerate shell as shown in Figure 3. The electronic correlations are most important for electrons occupying the degenerate shell and we will treat them using configuration-interaction (CI) method. The remaining background electrons are treated in HF. Hence we proceed to solve the HF problem for $97-7=90$ electrons on 97 sites with {$N_{\uparrow}=N_\downarrow=45$}. This leaves the degenerate shell at the Fermi level empty. We then rotate the many body Hamiltonian, Eq.\ref{eq:ManyBHam}, to the HF basis\cite{ozfidan2014microscopic,potasz2012electronic} for $N=90$ with final result given by
\begin{multline}
    \scalebox{1.0}{$H= \sum\limits_{p,\sigma}\epsilon^{HF}_{p\sigma}b_{p\sigma}^\dag b_{p\sigma}-\sum\limits_{p,q,\sigma} t_{pq\sigma}b_{p\sigma}^\dag b_{q\sigma}$}
    \\
    \scalebox{1.0}{$ + \frac{1}{2}\sum\limits_{p,q,r,s,\sigma,\sigma'}\braket{pq|V|rs}b_{p\sigma}^\dag b_{q\sigma'}^\dag b_{r\sigma'} b_{s\sigma} +\sum\limits_{p,q,\sigma} v^{add}_{pq}b_{p\sigma}^\dag b_{q\sigma}$}
       \label{eq:ManyBodyHInBasisofHF}
    \end{multline}
where
\begin{equation}
   t_{pq\sigma}=\sum\limits_{i,l}\tau_{il\sigma}a^*_{il\sigma}a_{lq\sigma}
\end{equation}
and
\begin{equation}
   b_{q\sigma}=\sum\limits_{l}a_{lq\sigma}c_{l\sigma}
\end{equation}
with {$\tau_{il\sigma}$} defined in Eqn.\ref{eq:tauil} but computed with respect to ATGQD density matrix elements. {$a_{lq\sigma}$} are the eigenvectors obtained by diagonalizing Eqn.\ref{eq:MFH}. We note that term {$t_{pq\sigma}$} appears, in order to lower the contribution of the quasiparticle-quasiparticle interaction term {~$\braket{pq|V|rs}$}. {$\braket{pq|V|rs}$} are Coulomb matrix elements in the basis of HF states, and are computed by rotating the real space matrix elements in Eq.\ref{eq:CoulombMatrixElements}, to the basis of HF states. They describe the remaining interaction of HF quasiparticles beyond the mean-field. The last term in Eq.\ref{eq:ManyBodyHInBasisofHF} involving  {$v_{pq}^{add}$} describes additional HF quasiparticles added to the degenerate shell. Since we solve the HF problem for {$N_e=90$} electrons, when adding HF quasiparticles at the CI level, we must compensate this charge with additional positive charges on the gate to maintain charge neutrality. Since a large gap separates the nearly degenerate shell from other states, it suffices to take only the shell near the Fermi level for CI calculations and neglect scattering from the valence band to the shell or from the shell to the conduction band.
Fig.\ref{fig:fig5} shows the low energy spectra obtained by diagonalizing Eq.\ref{eq:ManyBodyHInBasisofHF} for half-filled system with $N=97$ ,  with extra electron ($N=98$) and with extra hole ($N=96)$. Focusing on the half-filled (N=97 electrons) case in Fig.\ref{fig:fig5}(b),  we see the ground state of the half-filled shell to be maximally spin polarized, in agreement with Lieb’s theorem. The energy of this configuration is well separated from other states with lower total spin S, in other words the energy gap between our ground state, and first excited state with a different total spin $S$ is large. This implies that the energy cost to flip a spin is large. The removal or addition of a single electron, Fig.\ref{fig:fig5}(a) and Fig.\ref{fig:fig5}(c),results in a ground state which is still maximally spin polarized, but other low spin states lower their energy due to correlations, with  many total S states very close in energy. It costs practically zero energy to flip a spin in this case. In contrast to regular graphene \cite{potasz2012electronic} where the ground state corresponded to S=0, the ground state here has S=3, but we observe a dramatic drop of the energy gap between different total S states,  a phenomenon seen in graphene as well \cite{potasz2012electronic,gucclu2009magnetism,gucclu2011electric}. Correlations in the lower spin states cause a decrease of the energies, they become almost degenerate in energy with the maximum spin state. It is worth noting that the spin of the ground state for {$N=96$} electrons is in agreement with \cite{potasz2012electronic}, and the shrinking of the gap between different total spins states is consistent as well.
\subsection{ELECTRONIC CORRELATIONS IN THE ANTI-FERROMAGNETIC PHASE}
Due to the very different quasiparticle spectra for the AF phase and the semi-metallic phase as seen in Fig.\ref{fig:fig3}, we require two different approaches to  the many body problem. Here we solve the many body problem in the AF phase in real space by improving on the HF ground state in real space. In the AF phase there is no degenerate shell at the Fermi level. Hence, we begin  with the HF solution in the AF phase. We obtain the HF solution for a fixed number of spin up and spin down electrons by diagonalizing Eq.\ref{eq:MFH}, yielding a single Slater determinant defined as 
\begin{equation}
  \ket{HFGS} = \prod_{q=1}^{\lambda_F^\uparrow}b^{\dag}_{q\uparrow}\prod_{q=1}^{\lambda_F^\downarrow}b^{\dag}_{q\downarrow}\ket{0} ,
\end{equation}
where the Slater determinant is defined by filling up HF quasi particle levels up to the Fermi level for each spin. We then rotate this HF state to the site basis. In the site representation, we have a linear combination of Slater determinants, and we select the largest contributing state. For example, at half-filling for $S_z=\frac{7}{2}$, the ground state is given by
\begin{equation}
  \ket{GS} = \prod_{i\in A}c^{\dag}_{i\uparrow}\prod_{i\in B}c^{\dag}_{i\downarrow}\ket{0},
\end{equation}
where we place spin up electrons on the A sublattice, and spin down electrons on the B sublattice, representing a perfect antiferromagnetic phase. This is the largest dominant real space configuration composing that HF groundstate seen in Fig.\ref{fig:fig4}(b). We then divide the Hilbert space into segments for different {$S_z$} subspaces. This is done by starting with the ground state for different {$S_z$}, as shown above and constructing configurations with the same total {$S_z$}. The Hilbert space is divided into 5 sets of configurations defined by 
\begin{figure}
    \centering
    \includegraphics[width=1\linewidth]{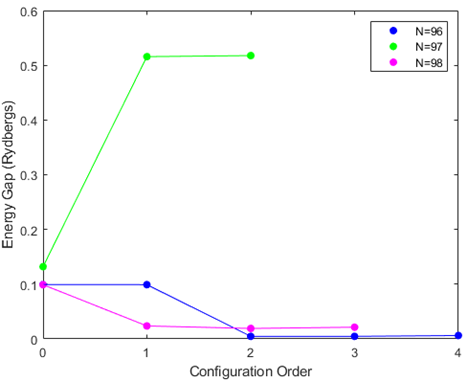}
    \caption{Energy Gap for different number of electrons as a function of increasing Hilbert space size. In green, we are at half-filling, and the gap between the ground state and the next spin state is very large, while for the case where we have added or removed an electron from the system, the energy gap collapses to almost zero.}
    \label{fig:fig6}
\end{figure}
\begin{figure}
    \centering
    \includegraphics[width=1\linewidth]{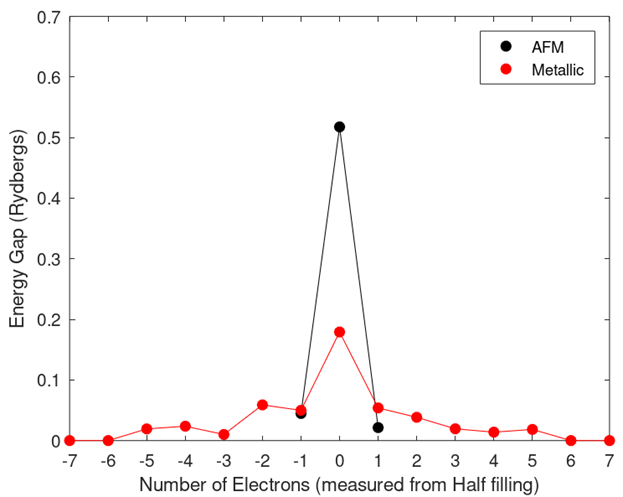}
    \caption{Energy gap vs filling factor, for AFM (black) and metallic regime (red). In both cases a similar behaviour is observed in which at half-filling the gap is large, and collapses away from half-filling.}
    \label{fig:fig7}
    \end{figure}
\begin{subequations}
\begin{equation}
\ket{O_{1,0}}=\sum\limits_{\substack{i,j,\sigma\\{<0DO>}}}c_{j\sigma} ^\dag c_{i\sigma} \ket{GS}
\end{equation}    
\begin{equation}
\ket{O_{1,1}}=\sum\limits_{\substack{i,j,\sigma\\{<1DO>}}}c_{j\sigma} ^\dag c_{i\sigma} \ket{GS}
\end{equation}
\begin{equation}
\ket{O_{2,0}}=\sum\limits_{\substack{i,j,k,l,\sigma,\sigma'\\{<0DO>}}}c_{l\sigma} ^\dag c_{k\sigma'} ^\dag c_{j\sigma'} c_{i\sigma} \ket{GS}
\end{equation}
\begin{equation}
\ket{O_{2,1}}=\sum\limits_{\substack{i,j,k,l,\sigma,\sigma'\\{<1DO>}}}c_{l\sigma} ^\dag c_{k\sigma'} ^\dag c_{j\sigma'} c_{i\sigma} \ket{GS} 
\end{equation}
\begin{equation}
\ket{O_{2,1}}=\sum\limits_{\substack{i,j,k,l,\sigma,\sigma',\\{<2DO>}}}c_{l\sigma} ^\dag c_{k\sigma'} ^\dag c_{j\sigma'} c_{i\sigma} \ket{GS},
\end{equation}
\end{subequations}
where the brackets under the sum denote a restriction of the configurations in which we include either 0,1 or 2 double occupancies (DO) measured from the number of DO of the ground state. The subscripts of $\ket{O_{\mu,\nu}}$ are defined with the first number denoting the number of electrons moved, and the second number being the number of extra double occupancies. We also restrict configurations by allowing only nearest neighbor scattering describing correlations. We then proceed to diagonalize the Hamiltonian in Eq.\ref{eq:ManyBHam}, as it is written in the site basis, and not in the HF basis. We take the same Coulomb matrix elements as described at the HF stage. After diagonalization, we observe a maximally spin polarized ground state in the case of $N=96, 97$ and $98$ electrons. In the case of adding or removing an electron, the energies are very close to each other. In Fig.\ref{fig:fig6} we observe that in the half-filled case we have a large gap separating this state from any other spin state. Meanwhile, when we add or remove an electron the energy gap between the spin polarized ground state, and the spin state closest in energy collapses to nearly zero. We now compare the metallic phase and AF phases. We first note that in the metallic phase, we see an emergence of a shell at zero energy of states which are practically degenerate. Here we are guided by the intuition that exchange interaction will lower the energy of a spin polarized half-filled system thus expecting Lieb’s theorem to be valid. In the AF regime, without the presence of a degenerate shell, we still see a spin polarized ground state. We then explore the regimes where we move away from half-filling, where Lieb’s theorem doesn't need to hold. In both regimes, we see a collapse of the energy gap. This is consistent with previous work \cite{potasz2012electronic,gucclu2009magnetism}. Fig.\ref{fig:fig7} shows schematically the collapse of the energy gap away from half-filling. The gap peaks at half-filling and collapses when we add/remove an electron.
We note that in the metallic regime, we study this as a function of filling factor, and see that the gap is maximum at half filling, where the spin polarized ground state is most stable.
\section{SUMMARY}
In summary, we described here how to construct a quantum simulator of an extended bipartite Hubbard model with broken sublattice symmetry using a structured lateral gate confining two dimensional electrons in a quantum well into artificial minima arranged in a hexagonal lattice. The sublattice symmetry breaking was generated by forming an artificial triangular graphene quantum dot (ATGQD) with zigzag edges. We demonstrated that in artificial graphene quantum dots, by tuning $U/t$, we can reach two distinct regimes, a semi-metallic and AFM. We showed for small systems that in both the metallic and AF regimes, the system at half-filling is partially spin polarized in agreement with Lieb’s theorem. The  addition or removal of an electron in both regimes collapses the energy gap and spin polarization. Such a simulator would allow simulation of larger systems, verification of results  presented here  and potential discovery of new phases resulting from strong electron-electron interactions in hexagonal lattice systems inherent in graphene and transition metal dichalcogenites.

\begin{acknowledgments}
Y.S., A.D.,M.C., M.K. and P.H. acknowledges support from 
NSERC QC2DM Strategic Grant No. STPG-521420, NSERC Discovery Grant No. RGPIN 2019-05714 and University of Ottawa Research Chair in
Quantum Theory of Quantum Materials, Nanostructures, and Devices.
Y.S. thanks J.Manalo, K.Sadecka, L.Szulakowska for useful discussions.
\end{acknowledgments}
 \appendix*
    \section{}
In this appendix we calculate the density matrix elements {$\rho_{jk\sigma'}^0$} in the AFM (see \cite{potasz2012electronic} for derivation in metallic phase). We start with the mean field Hamiltonian for bulk artificial graphene is given by 
\begin{multline}
    \scalebox{1.0}{$
    H^0_{MF}= \sum\limits_{i,l,\sigma}t_{il\sigma}c_{i\sigma}^\dag c_{l\sigma} $}
     \\
    \scalebox{1.0}{$ + \sum\limits_{i,j,k,l,\sigma,\sigma'} \left[\braket{ij|V|kl}-\braket{ij|V|lk}\delta_{\sigma \sigma'}\right]\rho^0_{jk\sigma'}c_{i\sigma}^\dag c_{l\sigma}$}
       \label{eq:MFHBulk}
    \end{multline}
Now, if we take only onsite density matrix elements, and we keep only terms where {$i=l$} (i.e. ignore small scattering elements) we can write the Hamiltonian as 
\begin{equation}
 H^0_{MF}=\sum\limits_{i,l,\sigma}t_{il}c_{i\sigma}^\dag c_{l\sigma}+\sum\limits_{i,\sigma}\Delta_{i\sigma}c_{i\sigma}^\dag c_{i\sigma}
       \label{eq:MFHBulkH}
\end{equation}   
Where
\begin{equation}
 \Delta_{i\sigma}=\sum\limits_{j\sigma'} \left[\braket{ij|V|ji}-\braket{ij|V|ij}\delta_{\sigma \sigma'}\right]\rho^0_{jj\sigma'}
       \label{eq:MFHBulkDiagonal}
\end{equation}   
Postulating that we are in the AFM regime, so that
\begin{subequations}
\begin{equation}
\rho_{jj\uparrow}^0=
\begin{cases}
      1 & j\in\text{Subblatice A}\\
      0 & j\in\text{Subblatice B}
    \end{cases}       
\end{equation}  
\begin{equation}
\rho_{jj\downarrow}^0=
\begin{cases}
      0 & j\in\text{Subblatice A}\\
      1 & j\in\text{Subblatice B}
    \end{cases}       
\end{equation}  
\end{subequations}
We arrive at the expressions 
\begin{subequations}
\begin{equation}
\Delta_{A\uparrow}=\sum\limits_{j>1}\braket{1j|V|j1}
\end{equation}  
\begin{equation}
\Delta_{A\downarrow}=U+\sum\limits_{j>1}\braket{1j|V|j1}
\end{equation}  
\begin{equation}
\Delta_{B\uparrow}=U+\sum\limits_{j>1}\braket{1j|V|j1}
\end{equation}  
\begin{equation}
\Delta_{B\downarrow}=\sum\limits_{j>1}\braket{1j|V|j1}
\end{equation}  
\end{subequations}
Where $\braket{1j|V|j1}$ corresponds to a direct interaction with a fixed site we call $i=1$, and another site j which is independent of which sublattice we fix our state $i=1$ to since all sites are identical. Now we can subtract the constant {$\frac{U}{2}+\sum\limits_{i>1}\braket{1i|V|i1}$} from all terms since this leads to only a shift in energies we get
\begin{subequations}
\begin{equation}
\Delta_{A\uparrow}=-\frac{U}{2}
\end{equation}  
\begin{equation}
\Delta_{A\downarrow}=\frac{U}{2}
\end{equation}  
\begin{equation}
\Delta_{B\uparrow}=\frac{U}{2}
\end{equation}  
\begin{equation}
\Delta_{B\downarrow}=-\frac{U}{2}
\end{equation}  
\end{subequations}
Taking nearest neighbours hopping only, the Hamiltonian in the basis of sublattices {$A\uparrow$},{$B\uparrow$},{$A\downarrow$},{$B\downarrow$} is given by
\begin{equation}
H\left(\vec{k}\right)=
  \begin{bmatrix}
    -\frac{U}{2} & -tf(\vec{k}) & 0 & 0 \\
   -tf^*(\vec{k}) & \frac{U}{2} & 0 &0 \\
    0 & 0 & \frac{U}{2}& -tf(\vec{k}) \\
    0 & 0 & -tf^*(\vec{k}) & -\frac{U}{2}
  \end{bmatrix}
\end{equation}  
This Hamiltonian is block diagonal, with each block mimicking gapped graphene. The valence band solution is given by
\begin{equation}
 E_k^-=-\sqrt{\left(\frac{U}{2}\right)^2+t^2|f(\vec{k})|^2}
\end{equation}  
\begin{subequations}
\begin{multline}
  \scalebox{1.0}{$\psi_k{_-}^{(\downarrow)}=-\sin\frac{\varphi_k}{2}e^{i\theta_k}\frac{1}{\sqrt{N_U}}\sum\limits_{\vec{R}_A}e^{i\vec{k}\cdot\vec{R}_A}\phi_z\left(\vec{r}-\vec{R}_A\right)$}
  \\
+ \scalebox{1.0}{$\cos\frac{\varphi_k}{2}\frac{1}{\sqrt{N_U}}\sum\limits_{\vec{R}_B}e^{i\vec{k}\cdot\vec{R}_B}\phi_z\left(\vec{r}-\vec{R}_B\right)$}
       \label{eq:GappedGrapheneEigenvectorDown}
    \end{multline}
 \begin{multline}
    \scalebox{1.0}{$\psi_{k-}^{(\uparrow)}=\cos\frac{\varphi_k}{2}\frac{1}{\sqrt{N_U}}\sum\limits_{\vec{R}_A}e^{i\vec{k}\cdot\vec{R}_A}\phi_z\left(\vec{r}-\vec{R}_A\right)$}
    \\
     \scalebox{1.0}{$-\sin\frac{\varphi_k}{2}\frac{1}{\sqrt{N_U}}\sum\limits_{\vec{R}_B}e^{i\vec{k}\cdot\vec{R}_B}\phi_z\left(\vec{r}-\vec{R}_B\right)$}
      \label{eq:GappedGrapheneEigenvectorUP}
     \end{multline}
\label{eq:GappedGrapheneEigenvector}
\end{subequations}
Where {$\frac{U}{2}=|E^-_k|\cos{\varphi_k}$} defines {$\varphi_k$} , and {$f(\vec{k})$} is the usual form factor of graphene. The density Matrix elements are defined as 
\begin{equation}
  \rho_{ij\sigma}^0=\sum_{\vec{k}}b^*_{\vec{R}_i,\vec{k},\sigma}b_{\vec{R}_j,\vec{k},\sigma}
\end{equation}  
Where {$b_{\vec{R}_j,\vec{k},\sigma}$} are the coeficients of the wavefunction defined in eqn.(\ref{eq:GappedGrapheneEigenvector}).  Computing the density matrix we have 
\begin{subequations}
\begin{equation}
\rho_{BB\downarrow}^0=\rho_{AA\uparrow}^0=\sum_{\vec{k}}\cos^2{\frac{\varphi}{2}}\approx 1
\end{equation}  
\begin{equation}
\rho_{BB\uparrow}^0=\rho_{AA\downarrow}^0=\sum_{\vec{k}}\left(1-\cos^2{\frac{\varphi}{2}}\right)\approx \frac{t}{U}
\end{equation}  
\begin{equation}
\rho_{BB\uparrow}^0=\rho_{AA\downarrow}^0=\sum_{\vec{k}}\sin{\frac{\varphi}{2}}\cos{\frac{\varphi}{2}}e^{-i\theta_k}e^{i\vec{k}\cdot{b}}\approx 0
\end{equation}  
\end{subequations}
Where the approximate solution exploits the fact that we are in the AF phase and so {$\frac{t}{U}<<1$}. This is in agreement with the numerical calculation of the density matrix elements in a finite flake in the center of the structure.

\newpage
\bibliographystyle{apsrev4-2}

\begin{thebibliography}{65}%
\makeatletter
\providecommand \@ifxundefined [1]{%
 \@ifx{#1\undefined}
}%
\providecommand \@ifnum [1]{%
 \ifnum #1\expandafter \@firstoftwo
 \else \expandafter \@secondoftwo
 \fi
}%
\providecommand \@ifx [1]{%
 \ifx #1\expandafter \@firstoftwo
 \else \expandafter \@secondoftwo
 \fi
}%
\providecommand \natexlab [1]{#1}%
\providecommand \enquote  [1]{``#1''}%
\providecommand \bibnamefont  [1]{#1}%
\providecommand \bibfnamefont [1]{#1}%
\providecommand \citenamefont [1]{#1}%
\providecommand \href@noop [0]{\@secondoftwo}%
\providecommand \href [0]{\begingroup \@sanitize@url \@href}%
\providecommand \@href[1]{\@@startlink{#1}\@@href}%
\providecommand \@@href[1]{\endgroup#1\@@endlink}%
\providecommand \@sanitize@url [0]{\catcode `\\12\catcode `\$12\catcode
  `\&12\catcode `\#12\catcode `\^12\catcode `\_12\catcode `\%12\relax}%
\providecommand \@@startlink[1]{}%
\providecommand \@@endlink[0]{}%
\providecommand \url  [0]{\begingroup\@sanitize@url \@url }%
\providecommand \@url [1]{\endgroup\@href {#1}{\urlprefix }}%
\providecommand \urlprefix  [0]{URL }%
\providecommand \Eprint [0]{\href }%
\providecommand \doibase [0]{https://doi.org/}%
\providecommand \selectlanguage [0]{\@gobble}%
\providecommand \bibinfo  [0]{\@secondoftwo}%
\providecommand \bibfield  [0]{\@secondoftwo}%
\providecommand \translation [1]{[#1]}%
\providecommand \BibitemOpen [0]{}%
\providecommand \bibitemStop [0]{}%
\providecommand \bibitemNoStop [0]{.\EOS\space}%
\providecommand \EOS [0]{\spacefactor3000\relax}%
\providecommand \BibitemShut  [1]{\csname bibitem#1\endcsname}%
\let\auto@bib@innerbib\@empty
\bibitem [{\citenamefont {Lloyd}(1996)}]{lloyd1996universal}%
  \BibitemOpen
  \bibfield  {author} {\bibinfo {author} {\bibfnamefont {S.}~\bibnamefont
  {Lloyd}},\ }\href@noop {} {\bibfield  {journal} {\bibinfo  {journal}
  {Science}\ ,\ \bibinfo {pages} {1073}} (\bibinfo {year} {1996})}\BibitemShut
  {NoStop}%
\bibitem [{\citenamefont {Bloch}(2005)}]{bloch2005ultracold}%
  \BibitemOpen
  \bibfield  {author} {\bibinfo {author} {\bibfnamefont {I.}~\bibnamefont
  {Bloch}},\ }\href@noop {} {\bibfield  {journal} {\bibinfo  {journal} {Nature
  physics}\ }\textbf {\bibinfo {volume} {1}},\ \bibinfo {pages} {23} (\bibinfo
  {year} {2005})}\BibitemShut {NoStop}%
\bibitem [{\citenamefont {Jaksch}\ and\ \citenamefont
  {Zoller}(2005)}]{jaksch2005cold}%
  \BibitemOpen
  \bibfield  {author} {\bibinfo {author} {\bibfnamefont {D.}~\bibnamefont
  {Jaksch}}\ and\ \bibinfo {author} {\bibfnamefont {P.}~\bibnamefont
  {Zoller}},\ }\href@noop {} {\bibfield  {journal} {\bibinfo  {journal} {Annals
  of physics}\ }\textbf {\bibinfo {volume} {315}},\ \bibinfo {pages} {52}
  (\bibinfo {year} {2005})}\BibitemShut {NoStop}%
\bibitem [{\citenamefont {Ortner}\ \emph {et~al.}(2009)\citenamefont {Ortner},
  \citenamefont {Micheli}, \citenamefont {Pupillo},\ and\ \citenamefont
  {Zoller}}]{ortner2009quantum}%
  \BibitemOpen
  \bibfield  {author} {\bibinfo {author} {\bibfnamefont {M.}~\bibnamefont
  {Ortner}}, \bibinfo {author} {\bibfnamefont {A.}~\bibnamefont {Micheli}},
  \bibinfo {author} {\bibfnamefont {G.}~\bibnamefont {Pupillo}},\ and\ \bibinfo
  {author} {\bibfnamefont {P.}~\bibnamefont {Zoller}},\ }\href@noop {}
  {\bibfield  {journal} {\bibinfo  {journal} {New Journal of Physics}\ }\textbf
  {\bibinfo {volume} {11}},\ \bibinfo {pages} {055045} (\bibinfo {year}
  {2009})}\BibitemShut {NoStop}%
\bibitem [{\citenamefont {Mazurenko}\ \emph {et~al.}(2017)\citenamefont
  {Mazurenko}, \citenamefont {Chiu}, \citenamefont {Ji}, \citenamefont
  {Parsons}, \citenamefont {Kan{\'a}sz-Nagy}, \citenamefont {Schmidt},
  \citenamefont {Grusdt}, \citenamefont {Demler}, \citenamefont {Greif},\ and\
  \citenamefont {Greiner}}]{mazurenko2017cold}%
  \BibitemOpen
  \bibfield  {author} {\bibinfo {author} {\bibfnamefont {A.}~\bibnamefont
  {Mazurenko}}, \bibinfo {author} {\bibfnamefont {C.~S.}\ \bibnamefont {Chiu}},
  \bibinfo {author} {\bibfnamefont {G.}~\bibnamefont {Ji}}, \bibinfo {author}
  {\bibfnamefont {M.~F.}\ \bibnamefont {Parsons}}, \bibinfo {author}
  {\bibfnamefont {M.}~\bibnamefont {Kan{\'a}sz-Nagy}}, \bibinfo {author}
  {\bibfnamefont {R.}~\bibnamefont {Schmidt}}, \bibinfo {author} {\bibfnamefont
  {F.}~\bibnamefont {Grusdt}}, \bibinfo {author} {\bibfnamefont
  {E.}~\bibnamefont {Demler}}, \bibinfo {author} {\bibfnamefont
  {D.}~\bibnamefont {Greif}},\ and\ \bibinfo {author} {\bibfnamefont
  {M.}~\bibnamefont {Greiner}},\ }\href@noop {} {\bibfield  {journal} {\bibinfo
   {journal} {Nature}\ }\textbf {\bibinfo {volume} {545}},\ \bibinfo {pages}
  {462} (\bibinfo {year} {2017})}\BibitemShut {NoStop}%
\bibitem [{\citenamefont {Kuhr}(2016)}]{kuhr2016quantum}%
  \BibitemOpen
  \bibfield  {author} {\bibinfo {author} {\bibfnamefont {S.}~\bibnamefont
  {Kuhr}},\ }\href@noop {} {\bibfield  {journal} {\bibinfo  {journal} {National
  Science Review}\ }\textbf {\bibinfo {volume} {3}},\ \bibinfo {pages} {170}
  (\bibinfo {year} {2016})}\BibitemShut {NoStop}%
\bibitem [{\citenamefont {Weimer}\ \emph {et~al.}(2010)\citenamefont {Weimer},
  \citenamefont {M{\"u}ller}, \citenamefont {Lesanovsky}, \citenamefont
  {Zoller},\ and\ \citenamefont {B{\"u}chler}}]{weimer2010rydberg}%
  \BibitemOpen
  \bibfield  {author} {\bibinfo {author} {\bibfnamefont {H.}~\bibnamefont
  {Weimer}}, \bibinfo {author} {\bibfnamefont {M.}~\bibnamefont {M{\"u}ller}},
  \bibinfo {author} {\bibfnamefont {I.}~\bibnamefont {Lesanovsky}}, \bibinfo
  {author} {\bibfnamefont {P.}~\bibnamefont {Zoller}},\ and\ \bibinfo {author}
  {\bibfnamefont {H.~P.}\ \bibnamefont {B{\"u}chler}},\ }\href@noop {}
  {\bibfield  {journal} {\bibinfo  {journal} {Nature Physics}\ }\textbf
  {\bibinfo {volume} {6}},\ \bibinfo {pages} {382} (\bibinfo {year}
  {2010})}\BibitemShut {NoStop}%
\bibitem [{\citenamefont {Barreiro}\ \emph {et~al.}(2011)\citenamefont
  {Barreiro}, \citenamefont {M{\"u}ller}, \citenamefont {Schindler},
  \citenamefont {Nigg}, \citenamefont {Monz}, \citenamefont {Chwalla},
  \citenamefont {Hennrich}, \citenamefont {Roos}, \citenamefont {Zoller},\ and\
  \citenamefont {Blatt}}]{barreiro2011open}%
  \BibitemOpen
  \bibfield  {author} {\bibinfo {author} {\bibfnamefont {J.~T.}\ \bibnamefont
  {Barreiro}}, \bibinfo {author} {\bibfnamefont {M.}~\bibnamefont
  {M{\"u}ller}}, \bibinfo {author} {\bibfnamefont {P.}~\bibnamefont
  {Schindler}}, \bibinfo {author} {\bibfnamefont {D.}~\bibnamefont {Nigg}},
  \bibinfo {author} {\bibfnamefont {T.}~\bibnamefont {Monz}}, \bibinfo {author}
  {\bibfnamefont {M.}~\bibnamefont {Chwalla}}, \bibinfo {author} {\bibfnamefont
  {M.}~\bibnamefont {Hennrich}}, \bibinfo {author} {\bibfnamefont {C.~F.}\
  \bibnamefont {Roos}}, \bibinfo {author} {\bibfnamefont {P.}~\bibnamefont
  {Zoller}},\ and\ \bibinfo {author} {\bibfnamefont {R.}~\bibnamefont
  {Blatt}},\ }\href@noop {} {\bibfield  {journal} {\bibinfo  {journal}
  {Nature}\ }\textbf {\bibinfo {volume} {470}},\ \bibinfo {pages} {486}
  (\bibinfo {year} {2011})}\BibitemShut {NoStop}%
\bibitem [{\citenamefont {Hempel}\ \emph {et~al.}(2018)\citenamefont {Hempel},
  \citenamefont {Maier}, \citenamefont {Romero}, \citenamefont {McClean},
  \citenamefont {Monz}, \citenamefont {Shen}, \citenamefont {Jurcevic},
  \citenamefont {Lanyon}, \citenamefont {Love}, \citenamefont {Babbush} \emph
  {et~al.}}]{hempel2018quantum}%
  \BibitemOpen
  \bibfield  {author} {\bibinfo {author} {\bibfnamefont {C.}~\bibnamefont
  {Hempel}}, \bibinfo {author} {\bibfnamefont {C.}~\bibnamefont {Maier}},
  \bibinfo {author} {\bibfnamefont {J.}~\bibnamefont {Romero}}, \bibinfo
  {author} {\bibfnamefont {J.}~\bibnamefont {McClean}}, \bibinfo {author}
  {\bibfnamefont {T.}~\bibnamefont {Monz}}, \bibinfo {author} {\bibfnamefont
  {H.}~\bibnamefont {Shen}}, \bibinfo {author} {\bibfnamefont {P.}~\bibnamefont
  {Jurcevic}}, \bibinfo {author} {\bibfnamefont {B.~P.}\ \bibnamefont
  {Lanyon}}, \bibinfo {author} {\bibfnamefont {P.}~\bibnamefont {Love}},
  \bibinfo {author} {\bibfnamefont {R.}~\bibnamefont {Babbush}}, \emph
  {et~al.},\ }\href@noop {} {\bibfield  {journal} {\bibinfo  {journal}
  {Physical Review X}\ }\textbf {\bibinfo {volume} {8}},\ \bibinfo {pages}
  {031022} (\bibinfo {year} {2018})}\BibitemShut {NoStop}%
\bibitem [{\citenamefont {Islam}\ \emph {et~al.}(2011)\citenamefont {Islam},
  \citenamefont {Edwards}, \citenamefont {Kim}, \citenamefont {Korenblit},
  \citenamefont {Noh}, \citenamefont {Carmichael}, \citenamefont {Lin},
  \citenamefont {Duan}, \citenamefont {Wang}, \citenamefont {Freericks} \emph
  {et~al.}}]{islam2011onset}%
  \BibitemOpen
  \bibfield  {author} {\bibinfo {author} {\bibfnamefont {R.}~\bibnamefont
  {Islam}}, \bibinfo {author} {\bibfnamefont {E.}~\bibnamefont {Edwards}},
  \bibinfo {author} {\bibfnamefont {K.}~\bibnamefont {Kim}}, \bibinfo {author}
  {\bibfnamefont {S.}~\bibnamefont {Korenblit}}, \bibinfo {author}
  {\bibfnamefont {C.}~\bibnamefont {Noh}}, \bibinfo {author} {\bibfnamefont
  {H.}~\bibnamefont {Carmichael}}, \bibinfo {author} {\bibfnamefont {G.-D.}\
  \bibnamefont {Lin}}, \bibinfo {author} {\bibfnamefont {L.-M.}\ \bibnamefont
  {Duan}}, \bibinfo {author} {\bibfnamefont {C.-C.~J.}\ \bibnamefont {Wang}},
  \bibinfo {author} {\bibfnamefont {J.}~\bibnamefont {Freericks}}, \emph
  {et~al.},\ }\href@noop {} {\bibfield  {journal} {\bibinfo  {journal} {Nature
  communications}\ }\textbf {\bibinfo {volume} {2}},\ \bibinfo {pages} {1}
  (\bibinfo {year} {2011})}\BibitemShut {NoStop}%
\bibitem [{\citenamefont {Leykam}\ \emph {et~al.}(2018)\citenamefont {Leykam},
  \citenamefont {Andreanov},\ and\ \citenamefont
  {Flach}}]{leykam2018artificial}%
  \BibitemOpen
  \bibfield  {author} {\bibinfo {author} {\bibfnamefont {D.}~\bibnamefont
  {Leykam}}, \bibinfo {author} {\bibfnamefont {A.}~\bibnamefont {Andreanov}},\
  and\ \bibinfo {author} {\bibfnamefont {S.}~\bibnamefont {Flach}},\
  }\href@noop {} {\bibfield  {journal} {\bibinfo  {journal} {Advances in
  Physics: X}\ }\textbf {\bibinfo {volume} {3}},\ \bibinfo {pages} {1473052}
  (\bibinfo {year} {2018})}\BibitemShut {NoStop}%
\bibitem [{\citenamefont {Salfi}\ \emph {et~al.}(2016)\citenamefont {Salfi},
  \citenamefont {Mol}, \citenamefont {Rahman}, \citenamefont {Klimeck},
  \citenamefont {Simmons}, \citenamefont {Hollenberg},\ and\ \citenamefont
  {Rogge}}]{salfi2016quantum}%
  \BibitemOpen
  \bibfield  {author} {\bibinfo {author} {\bibfnamefont {J.}~\bibnamefont
  {Salfi}}, \bibinfo {author} {\bibfnamefont {J.}~\bibnamefont {Mol}}, \bibinfo
  {author} {\bibfnamefont {R.}~\bibnamefont {Rahman}}, \bibinfo {author}
  {\bibfnamefont {G.}~\bibnamefont {Klimeck}}, \bibinfo {author} {\bibfnamefont
  {M.}~\bibnamefont {Simmons}}, \bibinfo {author} {\bibfnamefont
  {L.}~\bibnamefont {Hollenberg}},\ and\ \bibinfo {author} {\bibfnamefont
  {S.}~\bibnamefont {Rogge}},\ }\href@noop {} {\bibfield  {journal} {\bibinfo
  {journal} {Nature communications}\ }\textbf {\bibinfo {volume} {7}},\
  \bibinfo {pages} {1} (\bibinfo {year} {2016})}\BibitemShut {NoStop}%
\bibitem [{\citenamefont {Buluta}\ and\ \citenamefont
  {Nori}(2009)}]{buluta2009quantum}%
  \BibitemOpen
  \bibfield  {author} {\bibinfo {author} {\bibfnamefont {I.}~\bibnamefont
  {Buluta}}\ and\ \bibinfo {author} {\bibfnamefont {F.}~\bibnamefont {Nori}},\
  }\href@noop {} {\bibfield  {journal} {\bibinfo  {journal} {Science}\ }\textbf
  {\bibinfo {volume} {326}},\ \bibinfo {pages} {108} (\bibinfo {year}
  {2009})}\BibitemShut {NoStop}%
\bibitem [{\citenamefont {Aspuru-Guzik}\ and\ \citenamefont
  {Walther}(2012)}]{aspuru2012photonic}%
  \BibitemOpen
  \bibfield  {author} {\bibinfo {author} {\bibfnamefont {A.}~\bibnamefont
  {Aspuru-Guzik}}\ and\ \bibinfo {author} {\bibfnamefont {P.}~\bibnamefont
  {Walther}},\ }\href@noop {} {\bibfield  {journal} {\bibinfo  {journal}
  {Nature physics}\ }\textbf {\bibinfo {volume} {8}},\ \bibinfo {pages} {285}
  (\bibinfo {year} {2012})}\BibitemShut {NoStop}%
\bibitem [{\citenamefont {Bernien}\ \emph {et~al.}(2017)\citenamefont
  {Bernien}, \citenamefont {Schwartz}, \citenamefont {Keesling}, \citenamefont
  {Levine}, \citenamefont {Omran}, \citenamefont {Pichler}, \citenamefont
  {Choi}, \citenamefont {Zibrov}, \citenamefont {Endres}, \citenamefont
  {Greiner} \emph {et~al.}}]{bernien2017probing}%
  \BibitemOpen
  \bibfield  {author} {\bibinfo {author} {\bibfnamefont {H.}~\bibnamefont
  {Bernien}}, \bibinfo {author} {\bibfnamefont {S.}~\bibnamefont {Schwartz}},
  \bibinfo {author} {\bibfnamefont {A.}~\bibnamefont {Keesling}}, \bibinfo
  {author} {\bibfnamefont {H.}~\bibnamefont {Levine}}, \bibinfo {author}
  {\bibfnamefont {A.}~\bibnamefont {Omran}}, \bibinfo {author} {\bibfnamefont
  {H.}~\bibnamefont {Pichler}}, \bibinfo {author} {\bibfnamefont
  {S.}~\bibnamefont {Choi}}, \bibinfo {author} {\bibfnamefont {A.~S.}\
  \bibnamefont {Zibrov}}, \bibinfo {author} {\bibfnamefont {M.}~\bibnamefont
  {Endres}}, \bibinfo {author} {\bibfnamefont {M.}~\bibnamefont {Greiner}},
  \emph {et~al.},\ }\href@noop {} {\bibfield  {journal} {\bibinfo  {journal}
  {Nature}\ }\textbf {\bibinfo {volume} {551}},\ \bibinfo {pages} {579}
  (\bibinfo {year} {2017})}\BibitemShut {NoStop}%
\bibitem [{\citenamefont {Cai}\ \emph {et~al.}(2013)\citenamefont {Cai},
  \citenamefont {Retzker}, \citenamefont {Jelezko},\ and\ \citenamefont
  {Plenio}}]{cai2013large}%
  \BibitemOpen
  \bibfield  {author} {\bibinfo {author} {\bibfnamefont {J.}~\bibnamefont
  {Cai}}, \bibinfo {author} {\bibfnamefont {A.}~\bibnamefont {Retzker}},
  \bibinfo {author} {\bibfnamefont {F.}~\bibnamefont {Jelezko}},\ and\ \bibinfo
  {author} {\bibfnamefont {M.~B.}\ \bibnamefont {Plenio}},\ }\href@noop {}
  {\bibfield  {journal} {\bibinfo  {journal} {Nature Physics}\ }\textbf
  {\bibinfo {volume} {9}},\ \bibinfo {pages} {168} (\bibinfo {year}
  {2013})}\BibitemShut {NoStop}%
\bibitem [{\citenamefont {Dusko}\ \emph {et~al.}(2018)\citenamefont {Dusko},
  \citenamefont {Delgado}, \citenamefont {Saraiva},\ and\ \citenamefont
  {Koiller}}]{dusko2018adequacy}%
  \BibitemOpen
  \bibfield  {author} {\bibinfo {author} {\bibfnamefont {A.}~\bibnamefont
  {Dusko}}, \bibinfo {author} {\bibfnamefont {A.}~\bibnamefont {Delgado}},
  \bibinfo {author} {\bibfnamefont {A.}~\bibnamefont {Saraiva}},\ and\ \bibinfo
  {author} {\bibfnamefont {B.}~\bibnamefont {Koiller}},\ }\href@noop {}
  {\bibfield  {journal} {\bibinfo  {journal} {npj Quantum Information}\
  }\textbf {\bibinfo {volume} {4}},\ \bibinfo {pages} {1} (\bibinfo {year}
  {2018})}\BibitemShut {NoStop}%
\bibitem [{\citenamefont {Singha}\ \emph {et~al.}(2011)\citenamefont {Singha},
  \citenamefont {Gibertini}, \citenamefont {Karmakar}, \citenamefont {Yuan},
  \citenamefont {Polini}, \citenamefont {Vignale}, \citenamefont {Katsnelson},
  \citenamefont {Pinczuk}, \citenamefont {Pfeiffer}, \citenamefont {West} \emph
  {et~al.}}]{singha2011two}%
  \BibitemOpen
  \bibfield  {author} {\bibinfo {author} {\bibfnamefont {A.}~\bibnamefont
  {Singha}}, \bibinfo {author} {\bibfnamefont {M.}~\bibnamefont {Gibertini}},
  \bibinfo {author} {\bibfnamefont {B.}~\bibnamefont {Karmakar}}, \bibinfo
  {author} {\bibfnamefont {S.}~\bibnamefont {Yuan}}, \bibinfo {author}
  {\bibfnamefont {M.}~\bibnamefont {Polini}}, \bibinfo {author} {\bibfnamefont
  {G.}~\bibnamefont {Vignale}}, \bibinfo {author} {\bibfnamefont
  {M.}~\bibnamefont {Katsnelson}}, \bibinfo {author} {\bibfnamefont
  {A.}~\bibnamefont {Pinczuk}}, \bibinfo {author} {\bibfnamefont
  {L.}~\bibnamefont {Pfeiffer}}, \bibinfo {author} {\bibfnamefont
  {K.}~\bibnamefont {West}}, \emph {et~al.},\ }\href@noop {} {\bibfield
  {journal} {\bibinfo  {journal} {Science}\ }\textbf {\bibinfo {volume}
  {332}},\ \bibinfo {pages} {1176} (\bibinfo {year} {2011})}\BibitemShut
  {NoStop}%
\bibitem [{\citenamefont {Klembt}\ \emph {et~al.}(2018)\citenamefont {Klembt},
  \citenamefont {Harder}, \citenamefont {Egorov}, \citenamefont {K~Winkler},
  \citenamefont {Bandres}, \citenamefont {Emmerling}, \citenamefont
  {Worschech}, \citenamefont {Liew}, \citenamefont {Segev}, \citenamefont
  {Schneider},\ and\ \citenamefont {Höfling}}]{hofling2018}%
  \BibitemOpen
  \bibfield  {author} {\bibinfo {author} {\bibfnamefont {S.}~\bibnamefont
  {Klembt}}, \bibinfo {author} {\bibfnamefont {T.}~\bibnamefont {Harder}},
  \bibinfo {author} {\bibfnamefont {O.}~\bibnamefont {Egorov}}, \bibinfo
  {author} {\bibfnamefont {R.~G.}\ \bibnamefont {K~Winkler}}, \bibinfo {author}
  {\bibfnamefont {M.}~\bibnamefont {Bandres}}, \bibinfo {author} {\bibfnamefont
  {M.}~\bibnamefont {Emmerling}}, \bibinfo {author} {\bibfnamefont
  {L.}~\bibnamefont {Worschech}}, \bibinfo {author} {\bibfnamefont
  {T.}~\bibnamefont {Liew}}, \bibinfo {author} {\bibfnamefont {M.}~\bibnamefont
  {Segev}}, \bibinfo {author} {\bibfnamefont {C.}~\bibnamefont {Schneider}},\
  and\ \bibinfo {author} {\bibfnamefont {S.}~\bibnamefont {Höfling}},\
  }\href@noop {} {\bibfield  {journal} {\bibinfo  {journal} {Nature}\ }\textbf
  {\bibinfo {volume} {562}},\ \bibinfo {pages} {552} (\bibinfo {year}
  {2018})}\BibitemShut {NoStop}%
\bibitem [{\citenamefont {Park}\ and\ \citenamefont
  {Louie}(2009)}]{park2009making}%
  \BibitemOpen
  \bibfield  {author} {\bibinfo {author} {\bibfnamefont {C.-H.}\ \bibnamefont
  {Park}}\ and\ \bibinfo {author} {\bibfnamefont {S.~G.}\ \bibnamefont
  {Louie}},\ }\href@noop {} {\bibfield  {journal} {\bibinfo  {journal} {Nano
  letters}\ }\textbf {\bibinfo {volume} {9}},\ \bibinfo {pages} {1793}
  (\bibinfo {year} {2009})}\BibitemShut {NoStop}%
\bibitem [{\citenamefont {Li}\ \emph {et~al.}(2021)\citenamefont {Li},
  \citenamefont {Dietrich}, \citenamefont {Forsythe}, \citenamefont
  {Taniguchi}, \citenamefont {Watanabe}, \citenamefont {Moon},\ and\
  \citenamefont {Dean}}]{li2021anisotropic}%
  \BibitemOpen
  \bibfield  {author} {\bibinfo {author} {\bibfnamefont {Y.}~\bibnamefont
  {Li}}, \bibinfo {author} {\bibfnamefont {S.}~\bibnamefont {Dietrich}},
  \bibinfo {author} {\bibfnamefont {C.}~\bibnamefont {Forsythe}}, \bibinfo
  {author} {\bibfnamefont {T.}~\bibnamefont {Taniguchi}}, \bibinfo {author}
  {\bibfnamefont {K.}~\bibnamefont {Watanabe}}, \bibinfo {author}
  {\bibfnamefont {P.}~\bibnamefont {Moon}},\ and\ \bibinfo {author}
  {\bibfnamefont {C.~R.}\ \bibnamefont {Dean}},\ }\href@noop {} {\bibfield
  {journal} {\bibinfo  {journal} {Nature Nanotechnology}\ }\textbf {\bibinfo
  {volume} {16}},\ \bibinfo {pages} {525} (\bibinfo {year} {2021})}\BibitemShut
  {NoStop}%
\bibitem [{\citenamefont {Shi}\ \emph {et~al.}(2019)\citenamefont {Shi},
  \citenamefont {Ma},\ and\ \citenamefont {Song}}]{shi2019gate}%
  \BibitemOpen
  \bibfield  {author} {\bibinfo {author} {\bibfnamefont {L.-k.}\ \bibnamefont
  {Shi}}, \bibinfo {author} {\bibfnamefont {J.}~\bibnamefont {Ma}},\ and\
  \bibinfo {author} {\bibfnamefont {J.~C.}\ \bibnamefont {Song}},\ }\href@noop
  {} {\bibfield  {journal} {\bibinfo  {journal} {2D Materials}\ }\textbf
  {\bibinfo {volume} {7}},\ \bibinfo {pages} {015028} (\bibinfo {year}
  {2019})}\BibitemShut {NoStop}%
\bibitem [{\citenamefont {Forsythe}\ \emph {et~al.}(2018)\citenamefont
  {Forsythe}, \citenamefont {Zhou}, \citenamefont {Watanabe}, \citenamefont
  {Taniguchi}, \citenamefont {Pasupathy}, \citenamefont {Moon}, \citenamefont
  {Koshino}, \citenamefont {Kim},\ and\ \citenamefont
  {Dean}}]{forsythe2018band}%
  \BibitemOpen
  \bibfield  {author} {\bibinfo {author} {\bibfnamefont {C.}~\bibnamefont
  {Forsythe}}, \bibinfo {author} {\bibfnamefont {X.}~\bibnamefont {Zhou}},
  \bibinfo {author} {\bibfnamefont {K.}~\bibnamefont {Watanabe}}, \bibinfo
  {author} {\bibfnamefont {T.}~\bibnamefont {Taniguchi}}, \bibinfo {author}
  {\bibfnamefont {A.}~\bibnamefont {Pasupathy}}, \bibinfo {author}
  {\bibfnamefont {P.}~\bibnamefont {Moon}}, \bibinfo {author} {\bibfnamefont
  {M.}~\bibnamefont {Koshino}}, \bibinfo {author} {\bibfnamefont
  {P.}~\bibnamefont {Kim}},\ and\ \bibinfo {author} {\bibfnamefont {C.~R.}\
  \bibnamefont {Dean}},\ }\href@noop {} {\bibfield  {journal} {\bibinfo
  {journal} {Nature nanotechnology}\ }\textbf {\bibinfo {volume} {13}},\
  \bibinfo {pages} {566} (\bibinfo {year} {2018})}\BibitemShut {NoStop}%
\bibitem [{\citenamefont {Gibertini}\ \emph {et~al.}(2009)\citenamefont
  {Gibertini}, \citenamefont {Singha}, \citenamefont {Pellegrini},
  \citenamefont {Polini}, \citenamefont {Vignale}, \citenamefont {Pinczuk},
  \citenamefont {Pfeiffer},\ and\ \citenamefont
  {West}}]{gibertini2009engineering}%
  \BibitemOpen
  \bibfield  {author} {\bibinfo {author} {\bibfnamefont {M.}~\bibnamefont
  {Gibertini}}, \bibinfo {author} {\bibfnamefont {A.}~\bibnamefont {Singha}},
  \bibinfo {author} {\bibfnamefont {V.}~\bibnamefont {Pellegrini}}, \bibinfo
  {author} {\bibfnamefont {M.}~\bibnamefont {Polini}}, \bibinfo {author}
  {\bibfnamefont {G.}~\bibnamefont {Vignale}}, \bibinfo {author} {\bibfnamefont
  {A.}~\bibnamefont {Pinczuk}}, \bibinfo {author} {\bibfnamefont {L.~N.}\
  \bibnamefont {Pfeiffer}},\ and\ \bibinfo {author} {\bibfnamefont {K.~W.}\
  \bibnamefont {West}},\ }\href@noop {} {\bibfield  {journal} {\bibinfo
  {journal} {Physical Review B}\ }\textbf {\bibinfo {volume} {79}},\ \bibinfo
  {pages} {241406} (\bibinfo {year} {2009})}\BibitemShut {NoStop}%
\bibitem [{\citenamefont {Kyl{\"a}np{\"a}{\"a}}\ \emph
  {et~al.}(2016)\citenamefont {Kyl{\"a}np{\"a}{\"a}}, \citenamefont {Berardi},
  \citenamefont {R{\"a}s{\"a}nen}, \citenamefont {Garc{\'\i}a-Gonz{\'a}lez},
  \citenamefont {Rozzi},\ and\ \citenamefont {Rubio}}]{kylanpaa2016stability}%
  \BibitemOpen
  \bibfield  {author} {\bibinfo {author} {\bibfnamefont {I.}~\bibnamefont
  {Kyl{\"a}np{\"a}{\"a}}}, \bibinfo {author} {\bibfnamefont {F.}~\bibnamefont
  {Berardi}}, \bibinfo {author} {\bibfnamefont {E.}~\bibnamefont
  {R{\"a}s{\"a}nen}}, \bibinfo {author} {\bibfnamefont {P.}~\bibnamefont
  {Garc{\'\i}a-Gonz{\'a}lez}}, \bibinfo {author} {\bibfnamefont {C.~A.}\
  \bibnamefont {Rozzi}},\ and\ \bibinfo {author} {\bibfnamefont
  {A.}~\bibnamefont {Rubio}},\ }\href@noop {} {\bibfield  {journal} {\bibinfo
  {journal} {New Journal of Physics}\ }\textbf {\bibinfo {volume} {18}},\
  \bibinfo {pages} {083014} (\bibinfo {year} {2016})}\BibitemShut {NoStop}%
\bibitem [{\citenamefont {Uehlinger}\ \emph {et~al.}(2013)\citenamefont
  {Uehlinger}, \citenamefont {Jotzu}, \citenamefont {Messer}, \citenamefont
  {Greif}, \citenamefont {Hofstetter}, \citenamefont {Bissbort},\ and\
  \citenamefont {Esslinger}}]{uehlinger2013artificial}%
  \BibitemOpen
  \bibfield  {author} {\bibinfo {author} {\bibfnamefont {T.}~\bibnamefont
  {Uehlinger}}, \bibinfo {author} {\bibfnamefont {G.}~\bibnamefont {Jotzu}},
  \bibinfo {author} {\bibfnamefont {M.}~\bibnamefont {Messer}}, \bibinfo
  {author} {\bibfnamefont {D.}~\bibnamefont {Greif}}, \bibinfo {author}
  {\bibfnamefont {W.}~\bibnamefont {Hofstetter}}, \bibinfo {author}
  {\bibfnamefont {U.}~\bibnamefont {Bissbort}},\ and\ \bibinfo {author}
  {\bibfnamefont {T.}~\bibnamefont {Esslinger}},\ }\href@noop {} {\bibfield
  {journal} {\bibinfo  {journal} {Physical review letters}\ }\textbf {\bibinfo
  {volume} {111}},\ \bibinfo {pages} {185307} (\bibinfo {year}
  {2013})}\BibitemShut {NoStop}%
\bibitem [{\citenamefont {R{\"a}s{\"a}nen}\ \emph {et~al.}(2012)\citenamefont
  {R{\"a}s{\"a}nen}, \citenamefont {Rozzi}, \citenamefont {Pittalis},\ and\
  \citenamefont {Vignale}}]{rasanen2012electron}%
  \BibitemOpen
  \bibfield  {author} {\bibinfo {author} {\bibfnamefont {E.}~\bibnamefont
  {R{\"a}s{\"a}nen}}, \bibinfo {author} {\bibfnamefont {C.}~\bibnamefont
  {Rozzi}}, \bibinfo {author} {\bibfnamefont {S.}~\bibnamefont {Pittalis}},\
  and\ \bibinfo {author} {\bibfnamefont {G.}~\bibnamefont {Vignale}},\
  }\href@noop {} {\bibfield  {journal} {\bibinfo  {journal} {Physical review
  letters}\ }\textbf {\bibinfo {volume} {108}},\ \bibinfo {pages} {246803}
  (\bibinfo {year} {2012})}\BibitemShut {NoStop}%
\bibitem [{\citenamefont {Jacqmin}\ \emph {et~al.}(2014)\citenamefont
  {Jacqmin}, \citenamefont {Carusotto}, \citenamefont {Sagnes}, \citenamefont
  {Abbarchi}, \citenamefont {Solnyshkov}, \citenamefont {Malpuech},
  \citenamefont {Galopin}, \citenamefont {Lemaître}, \citenamefont {Bloch},\
  and\ \citenamefont {Amo}}]{bloch2014}%
  \BibitemOpen
  \bibfield  {author} {\bibinfo {author} {\bibfnamefont {T.}~\bibnamefont
  {Jacqmin}}, \bibinfo {author} {\bibfnamefont {I.}~\bibnamefont {Carusotto}},
  \bibinfo {author} {\bibfnamefont {I.}~\bibnamefont {Sagnes}}, \bibinfo
  {author} {\bibfnamefont {M.}~\bibnamefont {Abbarchi}}, \bibinfo {author}
  {\bibfnamefont {D.}~\bibnamefont {Solnyshkov}}, \bibinfo {author}
  {\bibfnamefont {G.}~\bibnamefont {Malpuech}}, \bibinfo {author}
  {\bibfnamefont {E.}~\bibnamefont {Galopin}}, \bibinfo {author} {\bibfnamefont
  {A.}~\bibnamefont {Lemaître}}, \bibinfo {author} {\bibfnamefont
  {J.}~\bibnamefont {Bloch}},\ and\ \bibinfo {author} {\bibfnamefont
  {A.}~\bibnamefont {Amo}},\ }\href@noop {} {\bibfield  {journal} {\bibinfo
  {journal} {Physical Review Letters}\ }\textbf {\bibinfo {volume} {112}},\
  \bibinfo {pages} {116402} (\bibinfo {year} {2014})}\BibitemShut {NoStop}%
\bibitem [{\citenamefont {Editors}\ \emph {et~al.}(2000)\citenamefont
  {Editors}, \citenamefont {EP2DS1999}, \citenamefont {Hawrylak}, \citenamefont
  {Lockwood},\ and\ \citenamefont {Sachrajda}}]{ep2ds1999}%
  \BibitemOpen
  \bibfield  {author} {\bibinfo {author} {\bibnamefont {Editors}}, \bibinfo
  {author} {\bibnamefont {EP2DS1999}}, \bibinfo {author} {\bibfnamefont
  {P.}~\bibnamefont {Hawrylak}}, \bibinfo {author} {\bibfnamefont
  {D.}~\bibnamefont {Lockwood}},\ and\ \bibinfo {author} {\bibfnamefont
  {A.}~\bibnamefont {Sachrajda}},\ }\href@noop {} {\bibfield  {journal}
  {\bibinfo  {journal} {Physica E}\ }\textbf {\bibinfo {volume} {6}} (\bibinfo
  {year} {2000})}\BibitemShut {NoStop}%
\bibitem [{\citenamefont {Editors}\ \emph {et~al.}(2010)\citenamefont
  {Editors}, \citenamefont {EP2DS2009}, \citenamefont {Katsumoto},
  \citenamefont {Kono},\ and\ \citenamefont {Tarucha}}]{ep2ds2009}%
  \BibitemOpen
  \bibfield  {author} {\bibinfo {author} {\bibnamefont {Editors}}, \bibinfo
  {author} {\bibnamefont {EP2DS2009}}, \bibinfo {author} {\bibfnamefont
  {S.}~\bibnamefont {Katsumoto}}, \bibinfo {author} {\bibfnamefont
  {K.}~\bibnamefont {Kono}},\ and\ \bibinfo {author} {\bibfnamefont
  {S.}~\bibnamefont {Tarucha}},\ }\href@noop {} {\bibfield  {journal} {\bibinfo
   {journal} {Physica E}\ }\textbf {\bibinfo {volume} {42}},\ \bibinfo {pages}
  {673} (\bibinfo {year} {2010})}\BibitemShut {NoStop}%
\bibitem [{\citenamefont {Piquero-Zulaica}\ \emph {et~al.}(2017)\citenamefont
  {Piquero-Zulaica}, \citenamefont {Lobo-Checa}, \citenamefont {Sadeghi},
  \citenamefont {Abd El-Fattah}, \citenamefont {Mitsui}, \citenamefont
  {Okamoto}, \citenamefont {Pawlak}, \citenamefont {Meier}, \citenamefont
  {Arnau}, \citenamefont {Ortega} \emph {et~al.}}]{piquero2017precise}%
  \BibitemOpen
  \bibfield  {author} {\bibinfo {author} {\bibfnamefont {I.}~\bibnamefont
  {Piquero-Zulaica}}, \bibinfo {author} {\bibfnamefont {J.}~\bibnamefont
  {Lobo-Checa}}, \bibinfo {author} {\bibfnamefont {A.}~\bibnamefont {Sadeghi}},
  \bibinfo {author} {\bibfnamefont {Z.~M.}\ \bibnamefont {Abd El-Fattah}},
  \bibinfo {author} {\bibfnamefont {C.}~\bibnamefont {Mitsui}}, \bibinfo
  {author} {\bibfnamefont {T.}~\bibnamefont {Okamoto}}, \bibinfo {author}
  {\bibfnamefont {R.}~\bibnamefont {Pawlak}}, \bibinfo {author} {\bibfnamefont
  {T.}~\bibnamefont {Meier}}, \bibinfo {author} {\bibfnamefont
  {A.}~\bibnamefont {Arnau}}, \bibinfo {author} {\bibfnamefont {J.~E.}\
  \bibnamefont {Ortega}}, \emph {et~al.},\ }\href@noop {} {\bibfield  {journal}
  {\bibinfo  {journal} {Nature communications}\ }\textbf {\bibinfo {volume}
  {8}},\ \bibinfo {pages} {1} (\bibinfo {year} {2017})}\BibitemShut {NoStop}%
\bibitem [{\citenamefont {Wang}\ \emph {et~al.}(2014)\citenamefont {Wang},
  \citenamefont {Tan}, \citenamefont {Wang}, \citenamefont {Louie},\ and\
  \citenamefont {Lin}}]{wang2014manipulation}%
  \BibitemOpen
  \bibfield  {author} {\bibinfo {author} {\bibfnamefont {S.}~\bibnamefont
  {Wang}}, \bibinfo {author} {\bibfnamefont {L.~Z.}\ \bibnamefont {Tan}},
  \bibinfo {author} {\bibfnamefont {W.}~\bibnamefont {Wang}}, \bibinfo {author}
  {\bibfnamefont {S.~G.}\ \bibnamefont {Louie}},\ and\ \bibinfo {author}
  {\bibfnamefont {N.}~\bibnamefont {Lin}},\ }\href@noop {} {\bibfield
  {journal} {\bibinfo  {journal} {Physical review letters}\ }\textbf {\bibinfo
  {volume} {113}},\ \bibinfo {pages} {196803} (\bibinfo {year}
  {2014})}\BibitemShut {NoStop}%
\bibitem [{\citenamefont {Wallace}(1947)}]{wallace1947band}%
  \BibitemOpen
  \bibfield  {author} {\bibinfo {author} {\bibfnamefont {P.~R.}\ \bibnamefont
  {Wallace}},\ }\href@noop {} {\bibfield  {journal} {\bibinfo  {journal}
  {Physical review}\ }\textbf {\bibinfo {volume} {71}},\ \bibinfo {pages} {622}
  (\bibinfo {year} {1947})}\BibitemShut {NoStop}%
\bibitem [{\citenamefont {Novoselov}\ \emph {et~al.}(2004)\citenamefont
  {Novoselov}, \citenamefont {Geim}, \citenamefont {Morozov}, \citenamefont
  {Jiang}, \citenamefont {Zhang}, \citenamefont {Dubonos}, \citenamefont
  {Grigorieva},\ and\ \citenamefont {Firsov}}]{novoselov2004electric}%
  \BibitemOpen
  \bibfield  {author} {\bibinfo {author} {\bibfnamefont {K.~S.}\ \bibnamefont
  {Novoselov}}, \bibinfo {author} {\bibfnamefont {A.~K.}\ \bibnamefont {Geim}},
  \bibinfo {author} {\bibfnamefont {S.~V.}\ \bibnamefont {Morozov}}, \bibinfo
  {author} {\bibfnamefont {D.-e.}\ \bibnamefont {Jiang}}, \bibinfo {author}
  {\bibfnamefont {Y.}~\bibnamefont {Zhang}}, \bibinfo {author} {\bibfnamefont
  {S.~V.}\ \bibnamefont {Dubonos}}, \bibinfo {author} {\bibfnamefont {I.~V.}\
  \bibnamefont {Grigorieva}},\ and\ \bibinfo {author} {\bibfnamefont {A.~A.}\
  \bibnamefont {Firsov}},\ }\href@noop {} {\bibfield  {journal} {\bibinfo
  {journal} {science}\ }\textbf {\bibinfo {volume} {306}},\ \bibinfo {pages}
  {666} (\bibinfo {year} {2004})}\BibitemShut {NoStop}%
\bibitem [{\citenamefont {Zhang}\ \emph {et~al.}(2005)\citenamefont {Zhang},
  \citenamefont {Tan}, \citenamefont {Stormer},\ and\ \citenamefont
  {Kim}}]{zhang2005experimental}%
  \BibitemOpen
  \bibfield  {author} {\bibinfo {author} {\bibfnamefont {Y.}~\bibnamefont
  {Zhang}}, \bibinfo {author} {\bibfnamefont {Y.-W.}\ \bibnamefont {Tan}},
  \bibinfo {author} {\bibfnamefont {H.~L.}\ \bibnamefont {Stormer}},\ and\
  \bibinfo {author} {\bibfnamefont {P.}~\bibnamefont {Kim}},\ }\href@noop {}
  {\bibfield  {journal} {\bibinfo  {journal} {nature}\ }\textbf {\bibinfo
  {volume} {438}},\ \bibinfo {pages} {201} (\bibinfo {year}
  {2005})}\BibitemShut {NoStop}%
\bibitem [{\citenamefont {Zhou}\ \emph {et~al.}(2006)\citenamefont {Zhou},
  \citenamefont {Gweon}, \citenamefont {Graf}, \citenamefont {Fedorov},
  \citenamefont {Spataru}, \citenamefont {Diehl}, \citenamefont {Kopelevich},
  \citenamefont {Lee}, \citenamefont {Louie},\ and\ \citenamefont
  {Lanzara}}]{zhou2006first}%
  \BibitemOpen
  \bibfield  {author} {\bibinfo {author} {\bibfnamefont {S.}~\bibnamefont
  {Zhou}}, \bibinfo {author} {\bibfnamefont {G.-H.}\ \bibnamefont {Gweon}},
  \bibinfo {author} {\bibfnamefont {J.}~\bibnamefont {Graf}}, \bibinfo {author}
  {\bibfnamefont {A.}~\bibnamefont {Fedorov}}, \bibinfo {author} {\bibfnamefont
  {C.}~\bibnamefont {Spataru}}, \bibinfo {author} {\bibfnamefont
  {R.}~\bibnamefont {Diehl}}, \bibinfo {author} {\bibfnamefont
  {Y.}~\bibnamefont {Kopelevich}}, \bibinfo {author} {\bibfnamefont {D.-H.}\
  \bibnamefont {Lee}}, \bibinfo {author} {\bibfnamefont {S.~G.}\ \bibnamefont
  {Louie}},\ and\ \bibinfo {author} {\bibfnamefont {A.}~\bibnamefont
  {Lanzara}},\ }\href@noop {} {\bibfield  {journal} {\bibinfo  {journal}
  {Nature physics}\ }\textbf {\bibinfo {volume} {2}},\ \bibinfo {pages} {595}
  (\bibinfo {year} {2006})}\BibitemShut {NoStop}%
\bibitem [{\citenamefont {Neto}\ \emph {et~al.}(2009)\citenamefont {Neto},
  \citenamefont {Guinea}, \citenamefont {Peres}, \citenamefont {Novoselov},\
  and\ \citenamefont {Geim}}]{neto2009electronic}%
  \BibitemOpen
  \bibfield  {author} {\bibinfo {author} {\bibfnamefont {A.~C.}\ \bibnamefont
  {Neto}}, \bibinfo {author} {\bibfnamefont {F.}~\bibnamefont {Guinea}},
  \bibinfo {author} {\bibfnamefont {N.~M.}\ \bibnamefont {Peres}}, \bibinfo
  {author} {\bibfnamefont {K.~S.}\ \bibnamefont {Novoselov}},\ and\ \bibinfo
  {author} {\bibfnamefont {A.~K.}\ \bibnamefont {Geim}},\ }\href@noop {}
  {\bibfield  {journal} {\bibinfo  {journal} {Reviews of modern physics}\
  }\textbf {\bibinfo {volume} {81}},\ \bibinfo {pages} {109} (\bibinfo {year}
  {2009})}\BibitemShut {NoStop}%
\bibitem [{\citenamefont {Novoselov}\ \emph {et~al.}(2005)\citenamefont
  {Novoselov}, \citenamefont {Geim}, \citenamefont {Morozov}, \citenamefont
  {Jiang}, \citenamefont {Katsnelson}, \citenamefont {Grigorieva},
  \citenamefont {Dubonos},\ and\ \citenamefont {Firsov}}]{novoselov2005two}%
  \BibitemOpen
  \bibfield  {author} {\bibinfo {author} {\bibfnamefont {K.~S.}\ \bibnamefont
  {Novoselov}}, \bibinfo {author} {\bibfnamefont {A.~K.}\ \bibnamefont {Geim}},
  \bibinfo {author} {\bibfnamefont {S.~V.}\ \bibnamefont {Morozov}}, \bibinfo
  {author} {\bibfnamefont {D.}~\bibnamefont {Jiang}}, \bibinfo {author}
  {\bibfnamefont {M.~I.}\ \bibnamefont {Katsnelson}}, \bibinfo {author}
  {\bibfnamefont {I.}~\bibnamefont {Grigorieva}}, \bibinfo {author}
  {\bibfnamefont {S.}~\bibnamefont {Dubonos}},\ and\ \bibinfo {author}
  {\bibfnamefont {A.}~\bibnamefont {Firsov}},\ }\href@noop {} {\bibfield
  {journal} {\bibinfo  {journal} {Nature}\ }\textbf {\bibinfo {volume} {438}},\
  \bibinfo {pages} {197} (\bibinfo {year} {2005})}\BibitemShut {NoStop}%
\bibitem [{\citenamefont {G{\"u}ttinger}\ \emph {et~al.}(2010)\citenamefont
  {G{\"u}ttinger}, \citenamefont {Frey}, \citenamefont {Stampfer},
  \citenamefont {Ihn},\ and\ \citenamefont {Ensslin}}]{guttinger2010spin}%
  \BibitemOpen
  \bibfield  {author} {\bibinfo {author} {\bibfnamefont {J.}~\bibnamefont
  {G{\"u}ttinger}}, \bibinfo {author} {\bibfnamefont {T.}~\bibnamefont {Frey}},
  \bibinfo {author} {\bibfnamefont {C.}~\bibnamefont {Stampfer}}, \bibinfo
  {author} {\bibfnamefont {T.}~\bibnamefont {Ihn}},\ and\ \bibinfo {author}
  {\bibfnamefont {K.}~\bibnamefont {Ensslin}},\ }\href@noop {} {\bibfield
  {journal} {\bibinfo  {journal} {Physical review letters}\ }\textbf {\bibinfo
  {volume} {105}},\ \bibinfo {pages} {116801} (\bibinfo {year}
  {2010})}\BibitemShut {NoStop}%
\bibitem [{\citenamefont {Saleem}\ \emph {et~al.}(2019)\citenamefont {Saleem},
  \citenamefont {Baldo}, \citenamefont {Delgado}, \citenamefont {Szulakowska},\
  and\ \citenamefont {Hawrylak}}]{saleem2019oscillations}%
  \BibitemOpen
  \bibfield  {author} {\bibinfo {author} {\bibfnamefont {Y.}~\bibnamefont
  {Saleem}}, \bibinfo {author} {\bibfnamefont {L.~N.}\ \bibnamefont {Baldo}},
  \bibinfo {author} {\bibfnamefont {A.}~\bibnamefont {Delgado}}, \bibinfo
  {author} {\bibfnamefont {L.}~\bibnamefont {Szulakowska}},\ and\ \bibinfo
  {author} {\bibfnamefont {P.}~\bibnamefont {Hawrylak}},\ }\href@noop {}
  {\bibfield  {journal} {\bibinfo  {journal} {Journal of Physics: Condensed
  Matter}\ }\textbf {\bibinfo {volume} {31}},\ \bibinfo {pages} {305503}
  (\bibinfo {year} {2019})}\BibitemShut {NoStop}%
\bibitem [{\citenamefont {Güçlü}\ \emph {et~al.}(2014)\citenamefont
  {Güçlü}, \citenamefont {Potasz}, \citenamefont {Korkusinski},\ and\
  \citenamefont {Hawrylak}}]{Guclu_graphene}%
  \BibitemOpen
  \bibfield  {author} {\bibinfo {author} {\bibfnamefont {A.~D.}\ \bibnamefont
  {Güçlü}}, \bibinfo {author} {\bibfnamefont {P.}~\bibnamefont {Potasz}},
  \bibinfo {author} {\bibfnamefont {M.}~\bibnamefont {Korkusinski}},\ and\
  \bibinfo {author} {\bibfnamefont {P.}~\bibnamefont {Hawrylak}},\ }\href@noop
  {} {\emph {\bibinfo {title} {Graphene Quantum Dots}}}\ (\bibinfo  {publisher}
  {Springer Verlag},\ \bibinfo {year} {2014})\BibitemShut {NoStop}%
\bibitem [{\citenamefont {G{\"u}{\c{c}}l{\"u}}\ \emph
  {et~al.}(2011)\citenamefont {G{\"u}{\c{c}}l{\"u}}, \citenamefont {Potasz},\
  and\ \citenamefont {Hawrylak}}]{gucclu2011electric}%
  \BibitemOpen
  \bibfield  {author} {\bibinfo {author} {\bibfnamefont {A.}~\bibnamefont
  {G{\"u}{\c{c}}l{\"u}}}, \bibinfo {author} {\bibfnamefont {P.}~\bibnamefont
  {Potasz}},\ and\ \bibinfo {author} {\bibfnamefont {P.}~\bibnamefont
  {Hawrylak}},\ }\href@noop {} {\bibfield  {journal} {\bibinfo  {journal}
  {Physical Review B}\ }\textbf {\bibinfo {volume} {84}},\ \bibinfo {pages}
  {035425} (\bibinfo {year} {2011})}\BibitemShut {NoStop}%
\bibitem [{\citenamefont {G{\"u}{\c{c}}l{\"u}}\ \emph
  {et~al.}(2009)\citenamefont {G{\"u}{\c{c}}l{\"u}}, \citenamefont {Potasz},
  \citenamefont {Voznyy}, \citenamefont {Korkusinski},\ and\ \citenamefont
  {Hawrylak}}]{gucclu2009magnetism}%
  \BibitemOpen
  \bibfield  {author} {\bibinfo {author} {\bibfnamefont {A.}~\bibnamefont
  {G{\"u}{\c{c}}l{\"u}}}, \bibinfo {author} {\bibfnamefont {P.}~\bibnamefont
  {Potasz}}, \bibinfo {author} {\bibfnamefont {O.}~\bibnamefont {Voznyy}},
  \bibinfo {author} {\bibfnamefont {M.}~\bibnamefont {Korkusinski}},\ and\
  \bibinfo {author} {\bibfnamefont {P.}~\bibnamefont {Hawrylak}},\ }\href@noop
  {} {\bibfield  {journal} {\bibinfo  {journal} {Physical review letters}\
  }\textbf {\bibinfo {volume} {103}},\ \bibinfo {pages} {246805} (\bibinfo
  {year} {2009})}\BibitemShut {NoStop}%
\bibitem [{\citenamefont {Potasz}\ \emph {et~al.}(2010)\citenamefont {Potasz},
  \citenamefont {G{\"u}{\c{c}}l{\"u}},\ and\ \citenamefont
  {Hawrylak}}]{potasz2010zero}%
  \BibitemOpen
  \bibfield  {author} {\bibinfo {author} {\bibfnamefont {P.}~\bibnamefont
  {Potasz}}, \bibinfo {author} {\bibfnamefont {A.}~\bibnamefont
  {G{\"u}{\c{c}}l{\"u}}},\ and\ \bibinfo {author} {\bibfnamefont
  {P.}~\bibnamefont {Hawrylak}},\ }\href@noop {} {\bibfield  {journal}
  {\bibinfo  {journal} {Physical Review B}\ }\textbf {\bibinfo {volume} {81}},\
  \bibinfo {pages} {033403} (\bibinfo {year} {2010})}\BibitemShut {NoStop}%
\bibitem [{\citenamefont {Potasz}\ \emph {et~al.}(2012)\citenamefont {Potasz},
  \citenamefont {G{\"u}{\c{c}}l{\"u}}, \citenamefont {W{\'o}js},\ and\
  \citenamefont {Hawrylak}}]{potasz2012electronic}%
  \BibitemOpen
  \bibfield  {author} {\bibinfo {author} {\bibfnamefont {P.}~\bibnamefont
  {Potasz}}, \bibinfo {author} {\bibfnamefont {A.}~\bibnamefont
  {G{\"u}{\c{c}}l{\"u}}}, \bibinfo {author} {\bibfnamefont {A.}~\bibnamefont
  {W{\'o}js}},\ and\ \bibinfo {author} {\bibfnamefont {P.}~\bibnamefont
  {Hawrylak}},\ }\href@noop {} {\bibfield  {journal} {\bibinfo  {journal}
  {Physical Review B}\ }\textbf {\bibinfo {volume} {85}},\ \bibinfo {pages}
  {075431} (\bibinfo {year} {2012})}\BibitemShut {NoStop}%
\bibitem [{\citenamefont {Ponomarenko}\ \emph {et~al.}(2008)\citenamefont
  {Ponomarenko}, \citenamefont {Schedin}, \citenamefont {Katsnelson},
  \citenamefont {Yang}, \citenamefont {Hill}, \citenamefont {Novoselov},\ and\
  \citenamefont {Geim}}]{ponomarenko2008chaotic}%
  \BibitemOpen
  \bibfield  {author} {\bibinfo {author} {\bibfnamefont {L.~A.}\ \bibnamefont
  {Ponomarenko}}, \bibinfo {author} {\bibfnamefont {F.}~\bibnamefont
  {Schedin}}, \bibinfo {author} {\bibfnamefont {M.~I.}\ \bibnamefont
  {Katsnelson}}, \bibinfo {author} {\bibfnamefont {R.}~\bibnamefont {Yang}},
  \bibinfo {author} {\bibfnamefont {E.~W.}\ \bibnamefont {Hill}}, \bibinfo
  {author} {\bibfnamefont {K.~S.}\ \bibnamefont {Novoselov}},\ and\ \bibinfo
  {author} {\bibfnamefont {A.~K.}\ \bibnamefont {Geim}},\ }\href@noop {}
  {\bibfield  {journal} {\bibinfo  {journal} {Science}\ }\textbf {\bibinfo
  {volume} {320}},\ \bibinfo {pages} {356} (\bibinfo {year}
  {2008})}\BibitemShut {NoStop}%
\bibitem [{\citenamefont {Lu}\ \emph {et~al.}(2011)\citenamefont {Lu},
  \citenamefont {Yeo}, \citenamefont {Gan}, \citenamefont {Wu},\ and\
  \citenamefont {Loh}}]{lu2011transforming}%
  \BibitemOpen
  \bibfield  {author} {\bibinfo {author} {\bibfnamefont {J.}~\bibnamefont
  {Lu}}, \bibinfo {author} {\bibfnamefont {P.~S.~E.}\ \bibnamefont {Yeo}},
  \bibinfo {author} {\bibfnamefont {C.~K.}\ \bibnamefont {Gan}}, \bibinfo
  {author} {\bibfnamefont {P.}~\bibnamefont {Wu}},\ and\ \bibinfo {author}
  {\bibfnamefont {K.~P.}\ \bibnamefont {Loh}},\ }\href@noop {} {\bibfield
  {journal} {\bibinfo  {journal} {Nature nanotechnology}\ }\textbf {\bibinfo
  {volume} {6}},\ \bibinfo {pages} {247} (\bibinfo {year} {2011})}\BibitemShut
  {NoStop}%
\bibitem [{\citenamefont {Ezawa}(2010)}]{ezawa2010dirac}%
  \BibitemOpen
  \bibfield  {author} {\bibinfo {author} {\bibfnamefont {M.}~\bibnamefont
  {Ezawa}},\ }\href@noop {} {\bibfield  {journal} {\bibinfo  {journal}
  {Physical Review B}\ }\textbf {\bibinfo {volume} {81}},\ \bibinfo {pages}
  {201402} (\bibinfo {year} {2010})}\BibitemShut {NoStop}%
\bibitem [{\citenamefont {Wunsch}\ \emph {et~al.}(2008)\citenamefont {Wunsch},
  \citenamefont {Stauber},\ and\ \citenamefont {Guinea}}]{wunsch2008electron}%
  \BibitemOpen
  \bibfield  {author} {\bibinfo {author} {\bibfnamefont {B.}~\bibnamefont
  {Wunsch}}, \bibinfo {author} {\bibfnamefont {T.}~\bibnamefont {Stauber}},\
  and\ \bibinfo {author} {\bibfnamefont {F.}~\bibnamefont {Guinea}},\
  }\href@noop {} {\bibfield  {journal} {\bibinfo  {journal} {Physical Review
  B}\ }\textbf {\bibinfo {volume} {77}},\ \bibinfo {pages} {035316} (\bibinfo
  {year} {2008})}\BibitemShut {NoStop}%
\bibitem [{\citenamefont {Peleg}\ \emph {et~al.}(2007)\citenamefont {Peleg},
  \citenamefont {Bartal}, \citenamefont {Freedman}, \citenamefont {Manela},
  \citenamefont {Segev},\ and\ \citenamefont
  {Christodoulides}}]{peleg2007conical}%
  \BibitemOpen
  \bibfield  {author} {\bibinfo {author} {\bibfnamefont {O.}~\bibnamefont
  {Peleg}}, \bibinfo {author} {\bibfnamefont {G.}~\bibnamefont {Bartal}},
  \bibinfo {author} {\bibfnamefont {B.}~\bibnamefont {Freedman}}, \bibinfo
  {author} {\bibfnamefont {O.}~\bibnamefont {Manela}}, \bibinfo {author}
  {\bibfnamefont {M.}~\bibnamefont {Segev}},\ and\ \bibinfo {author}
  {\bibfnamefont {D.~N.}\ \bibnamefont {Christodoulides}},\ }\href@noop {}
  {\bibfield  {journal} {\bibinfo  {journal} {Physical review letters}\
  }\textbf {\bibinfo {volume} {98}},\ \bibinfo {pages} {103901} (\bibinfo
  {year} {2007})}\BibitemShut {NoStop}%
\bibitem [{\citenamefont {Gomes}\ \emph {et~al.}(2012)\citenamefont {Gomes},
  \citenamefont {Mar}, \citenamefont {Ko}, \citenamefont {Guinea},\ and\
  \citenamefont {Manoharan}}]{gomes2012designer}%
  \BibitemOpen
  \bibfield  {author} {\bibinfo {author} {\bibfnamefont {K.~K.}\ \bibnamefont
  {Gomes}}, \bibinfo {author} {\bibfnamefont {W.}~\bibnamefont {Mar}}, \bibinfo
  {author} {\bibfnamefont {W.}~\bibnamefont {Ko}}, \bibinfo {author}
  {\bibfnamefont {F.}~\bibnamefont {Guinea}},\ and\ \bibinfo {author}
  {\bibfnamefont {H.~C.}\ \bibnamefont {Manoharan}},\ }\href@noop {} {\bibfield
   {journal} {\bibinfo  {journal} {Nature}\ }\textbf {\bibinfo {volume}
  {483}},\ \bibinfo {pages} {306} (\bibinfo {year} {2012})}\BibitemShut
  {NoStop}%
\bibitem [{\citenamefont {Drost}\ \emph {et~al.}(2017)\citenamefont {Drost},
  \citenamefont {Ojanen}, \citenamefont {Harju},\ and\ \citenamefont
  {Liljeroth}}]{drost2017topological}%
  \BibitemOpen
  \bibfield  {author} {\bibinfo {author} {\bibfnamefont {R.}~\bibnamefont
  {Drost}}, \bibinfo {author} {\bibfnamefont {T.}~\bibnamefont {Ojanen}},
  \bibinfo {author} {\bibfnamefont {A.}~\bibnamefont {Harju}},\ and\ \bibinfo
  {author} {\bibfnamefont {P.}~\bibnamefont {Liljeroth}},\ }\href@noop {}
  {\bibfield  {journal} {\bibinfo  {journal} {Nature Physics}\ }\textbf
  {\bibinfo {volume} {13}},\ \bibinfo {pages} {668} (\bibinfo {year}
  {2017})}\BibitemShut {NoStop}%
\bibitem [{\citenamefont {Slot}\ \emph {et~al.}(2017)\citenamefont {Slot},
  \citenamefont {Gardenier}, \citenamefont {Jacobse}, \citenamefont {van
  Miert}, \citenamefont {Kempkes}, \citenamefont {Zevenhuizen}, \citenamefont
  {Smith}, \citenamefont {Vanmaekelbergh},\ and\ \citenamefont
  {Swart}}]{slot2017experimental}%
  \BibitemOpen
  \bibfield  {author} {\bibinfo {author} {\bibfnamefont {M.~R.}\ \bibnamefont
  {Slot}}, \bibinfo {author} {\bibfnamefont {T.~S.}\ \bibnamefont {Gardenier}},
  \bibinfo {author} {\bibfnamefont {P.~H.}\ \bibnamefont {Jacobse}}, \bibinfo
  {author} {\bibfnamefont {G.~C.}\ \bibnamefont {van Miert}}, \bibinfo {author}
  {\bibfnamefont {S.~N.}\ \bibnamefont {Kempkes}}, \bibinfo {author}
  {\bibfnamefont {S.~J.}\ \bibnamefont {Zevenhuizen}}, \bibinfo {author}
  {\bibfnamefont {C.~M.}\ \bibnamefont {Smith}}, \bibinfo {author}
  {\bibfnamefont {D.}~\bibnamefont {Vanmaekelbergh}},\ and\ \bibinfo {author}
  {\bibfnamefont {I.}~\bibnamefont {Swart}},\ }\href@noop {} {\bibfield
  {journal} {\bibinfo  {journal} {Nature physics}\ }\textbf {\bibinfo {volume}
  {13}},\ \bibinfo {pages} {672} (\bibinfo {year} {2017})}\BibitemShut
  {NoStop}%
\bibitem [{\citenamefont {Lieb}(1989)}]{lieb1989two}%
  \BibitemOpen
  \bibfield  {author} {\bibinfo {author} {\bibfnamefont {E.~H.}\ \bibnamefont
  {Lieb}},\ }\href@noop {} {\bibfield  {journal} {\bibinfo  {journal} {Physical
  review letters}\ }\textbf {\bibinfo {volume} {62}},\ \bibinfo {pages} {1201}
  (\bibinfo {year} {1989})}\BibitemShut {NoStop}%
\bibitem [{\citenamefont {Ezawa}(2007)}]{ezawa2007metallic}%
  \BibitemOpen
  \bibfield  {author} {\bibinfo {author} {\bibfnamefont {M.}~\bibnamefont
  {Ezawa}},\ }\href@noop {} {\bibfield  {journal} {\bibinfo  {journal}
  {Physical Review B}\ }\textbf {\bibinfo {volume} {76}},\ \bibinfo {pages}
  {245415} (\bibinfo {year} {2007})}\BibitemShut {NoStop}%
\bibitem [{\citenamefont {Fern{\'a}ndez-Rossier}\ and\ \citenamefont
  {Palacios}(2007)}]{fernandez2007magnetism}%
  \BibitemOpen
  \bibfield  {author} {\bibinfo {author} {\bibfnamefont {J.}~\bibnamefont
  {Fern{\'a}ndez-Rossier}}\ and\ \bibinfo {author} {\bibfnamefont {J.~J.}\
  \bibnamefont {Palacios}},\ }\href@noop {} {\bibfield  {journal} {\bibinfo
  {journal} {Physical Review Letters}\ }\textbf {\bibinfo {volume} {99}},\
  \bibinfo {pages} {177204} (\bibinfo {year} {2007})}\BibitemShut {NoStop}%
\bibitem [{\citenamefont {Wang}\ \emph {et~al.}(2009)\citenamefont {Wang},
  \citenamefont {Yazyev}, \citenamefont {Meng},\ and\ \citenamefont
  {Kaxiras}}]{wang2009topological}%
  \BibitemOpen
  \bibfield  {author} {\bibinfo {author} {\bibfnamefont {W.~L.}\ \bibnamefont
  {Wang}}, \bibinfo {author} {\bibfnamefont {O.~V.}\ \bibnamefont {Yazyev}},
  \bibinfo {author} {\bibfnamefont {S.}~\bibnamefont {Meng}},\ and\ \bibinfo
  {author} {\bibfnamefont {E.}~\bibnamefont {Kaxiras}},\ }\href@noop {}
  {\bibfield  {journal} {\bibinfo  {journal} {Physical review letters}\
  }\textbf {\bibinfo {volume} {102}},\ \bibinfo {pages} {157201} (\bibinfo
  {year} {2009})}\BibitemShut {NoStop}%
\bibitem [{\citenamefont {Pavli{\v{c}}ek}\ \emph {et~al.}(2017)\citenamefont
  {Pavli{\v{c}}ek}, \citenamefont {Mistry}, \citenamefont {Majzik},
  \citenamefont {Moll}, \citenamefont {Meyer}, \citenamefont {Fox},\ and\
  \citenamefont {Gross}}]{pavlivcek2017synthesis}%
  \BibitemOpen
  \bibfield  {author} {\bibinfo {author} {\bibfnamefont {N.}~\bibnamefont
  {Pavli{\v{c}}ek}}, \bibinfo {author} {\bibfnamefont {A.}~\bibnamefont
  {Mistry}}, \bibinfo {author} {\bibfnamefont {Z.}~\bibnamefont {Majzik}},
  \bibinfo {author} {\bibfnamefont {N.}~\bibnamefont {Moll}}, \bibinfo {author}
  {\bibfnamefont {G.}~\bibnamefont {Meyer}}, \bibinfo {author} {\bibfnamefont
  {D.~J.}\ \bibnamefont {Fox}},\ and\ \bibinfo {author} {\bibfnamefont
  {L.}~\bibnamefont {Gross}},\ }\href@noop {} {\bibfield  {journal} {\bibinfo
  {journal} {Nature nanotechnology}\ }\textbf {\bibinfo {volume} {12}},\
  \bibinfo {pages} {308} (\bibinfo {year} {2017})}\BibitemShut {NoStop}%
\bibitem [{\citenamefont {Mishra}\ \emph {et~al.}(2020)\citenamefont {Mishra},
  \citenamefont {Beyer}, \citenamefont {Eimre}, \citenamefont {Kezilebieke},
  \citenamefont {Berger}, \citenamefont {Gr{\"o}ning}, \citenamefont
  {Pignedoli}, \citenamefont {M{\"u}llen}, \citenamefont {Liljeroth},
  \citenamefont {Ruffieux} \emph {et~al.}}]{mishra2020topological}%
  \BibitemOpen
  \bibfield  {author} {\bibinfo {author} {\bibfnamefont {S.}~\bibnamefont
  {Mishra}}, \bibinfo {author} {\bibfnamefont {D.}~\bibnamefont {Beyer}},
  \bibinfo {author} {\bibfnamefont {K.}~\bibnamefont {Eimre}}, \bibinfo
  {author} {\bibfnamefont {S.}~\bibnamefont {Kezilebieke}}, \bibinfo {author}
  {\bibfnamefont {R.}~\bibnamefont {Berger}}, \bibinfo {author} {\bibfnamefont
  {O.}~\bibnamefont {Gr{\"o}ning}}, \bibinfo {author} {\bibfnamefont {C.~A.}\
  \bibnamefont {Pignedoli}}, \bibinfo {author} {\bibfnamefont {K.}~\bibnamefont
  {M{\"u}llen}}, \bibinfo {author} {\bibfnamefont {P.}~\bibnamefont
  {Liljeroth}}, \bibinfo {author} {\bibfnamefont {P.}~\bibnamefont {Ruffieux}},
  \emph {et~al.},\ }\href@noop {} {\bibfield  {journal} {\bibinfo  {journal}
  {Nature nanotechnology}\ }\textbf {\bibinfo {volume} {15}},\ \bibinfo {pages}
  {22} (\bibinfo {year} {2020})}\BibitemShut {NoStop}%
\bibitem [{\citenamefont {Su}\ \emph {et~al.}(2019)\citenamefont {Su},
  \citenamefont {Telychko}, \citenamefont {Hu}, \citenamefont {Macam},
  \citenamefont {Mutombo}, \citenamefont {Zhang}, \citenamefont {Bao},
  \citenamefont {Cheng}, \citenamefont {Huang}, \citenamefont {Qiu} \emph
  {et~al.}}]{su2019atomically}%
  \BibitemOpen
  \bibfield  {author} {\bibinfo {author} {\bibfnamefont {J.}~\bibnamefont
  {Su}}, \bibinfo {author} {\bibfnamefont {M.}~\bibnamefont {Telychko}},
  \bibinfo {author} {\bibfnamefont {P.}~\bibnamefont {Hu}}, \bibinfo {author}
  {\bibfnamefont {G.}~\bibnamefont {Macam}}, \bibinfo {author} {\bibfnamefont
  {P.}~\bibnamefont {Mutombo}}, \bibinfo {author} {\bibfnamefont
  {H.}~\bibnamefont {Zhang}}, \bibinfo {author} {\bibfnamefont
  {Y.}~\bibnamefont {Bao}}, \bibinfo {author} {\bibfnamefont {F.}~\bibnamefont
  {Cheng}}, \bibinfo {author} {\bibfnamefont {Z.-Q.}\ \bibnamefont {Huang}},
  \bibinfo {author} {\bibfnamefont {Z.}~\bibnamefont {Qiu}}, \emph {et~al.},\
  }\href@noop {} {\bibfield  {journal} {\bibinfo  {journal} {Science advances}\
  }\textbf {\bibinfo {volume} {5}},\ \bibinfo {pages} {eaav7717} (\bibinfo
  {year} {2019})}\BibitemShut {NoStop}%
\bibitem [{\citenamefont {Wehling}\ \emph {et~al.}(2011)\citenamefont
  {Wehling}, \citenamefont {Şaşıoğlu}, \citenamefont {Friedrich},
  \citenamefont {Lichtenstein}, \citenamefont {Katsnelson},\ and\ \citenamefont
  {Blügel}}]{wehling2011}%
  \BibitemOpen
  \bibfield  {author} {\bibinfo {author} {\bibfnamefont {T.}~\bibnamefont
  {Wehling}}, \bibinfo {author} {\bibfnamefont {E.}~\bibnamefont
  {Şaşıoğlu}}, \bibinfo {author} {\bibfnamefont {C.}~\bibnamefont
  {Friedrich}}, \bibinfo {author} {\bibfnamefont {A.}~\bibnamefont
  {Lichtenstein}}, \bibinfo {author} {\bibfnamefont {M.}~\bibnamefont
  {Katsnelson}},\ and\ \bibinfo {author} {\bibfnamefont {S.}~\bibnamefont
  {Blügel}},\ }\href@noop {} {\bibfield  {journal} {\bibinfo  {journal}
  {Physical Review Letters}\ }\textbf {\bibinfo {volume} {106}},\ \bibinfo
  {pages} {236805} (\bibinfo {year} {2011})}\BibitemShut {NoStop}%
\bibitem [{\citenamefont {Wehling}\ \emph {et~al.}(2014)\citenamefont
  {Wehling}, \citenamefont {Black-Schaffer},\ and\ \citenamefont
  {Balatsky}}]{wehling2014dirac}%
  \BibitemOpen
  \bibfield  {author} {\bibinfo {author} {\bibfnamefont {T.}~\bibnamefont
  {Wehling}}, \bibinfo {author} {\bibfnamefont {A.~M.}\ \bibnamefont
  {Black-Schaffer}},\ and\ \bibinfo {author} {\bibfnamefont {A.~V.}\
  \bibnamefont {Balatsky}},\ }\href@noop {} {\bibfield  {journal} {\bibinfo
  {journal} {Advances in Physics}\ }\textbf {\bibinfo {volume} {63}},\ \bibinfo
  {pages} {1} (\bibinfo {year} {2014})}\BibitemShut {NoStop}%
\bibitem [{\citenamefont {Sorella}\ and\ \citenamefont
  {Tosatti}(1992)}]{sorella1992semi}%
  \BibitemOpen
  \bibfield  {author} {\bibinfo {author} {\bibfnamefont {S.}~\bibnamefont
  {Sorella}}\ and\ \bibinfo {author} {\bibfnamefont {E.}~\bibnamefont
  {Tosatti}},\ }\href@noop {} {\bibfield  {journal} {\bibinfo  {journal} {EPL
  (Europhysics Letters)}\ }\textbf {\bibinfo {volume} {19}},\ \bibinfo {pages}
  {699} (\bibinfo {year} {1992})}\BibitemShut {NoStop}%
\bibitem [{\citenamefont {Voznyy}\ \emph {et~al.}(2011)\citenamefont {Voznyy},
  \citenamefont {G{\"u}{\c{c}}l{\"u}}, \citenamefont {Potasz},\ and\
  \citenamefont {Hawrylak}}]{voznyy2011effect}%
  \BibitemOpen
  \bibfield  {author} {\bibinfo {author} {\bibfnamefont {O.}~\bibnamefont
  {Voznyy}}, \bibinfo {author} {\bibfnamefont {A.~D.}\ \bibnamefont
  {G{\"u}{\c{c}}l{\"u}}}, \bibinfo {author} {\bibfnamefont {P.}~\bibnamefont
  {Potasz}},\ and\ \bibinfo {author} {\bibfnamefont {P.}~\bibnamefont
  {Hawrylak}},\ }\href@noop {} {\bibfield  {journal} {\bibinfo  {journal}
  {Physical Review B}\ }\textbf {\bibinfo {volume} {83}},\ \bibinfo {pages}
  {165417} (\bibinfo {year} {2011})}\BibitemShut {NoStop}%
\bibitem [{\citenamefont {Ozfidan}\ \emph {et~al.}(2014)\citenamefont
  {Ozfidan}, \citenamefont {Korkusinski}, \citenamefont {G{\"u}{\c{c}}l{\"u}},
  \citenamefont {McGuire},\ and\ \citenamefont
  {Hawrylak}}]{ozfidan2014microscopic}%
  \BibitemOpen
  \bibfield  {author} {\bibinfo {author} {\bibfnamefont {I.}~\bibnamefont
  {Ozfidan}}, \bibinfo {author} {\bibfnamefont {M.}~\bibnamefont
  {Korkusinski}}, \bibinfo {author} {\bibfnamefont {A.~D.}\ \bibnamefont
  {G{\"u}{\c{c}}l{\"u}}}, \bibinfo {author} {\bibfnamefont {J.~A.}\
  \bibnamefont {McGuire}},\ and\ \bibinfo {author} {\bibfnamefont
  {P.}~\bibnamefont {Hawrylak}},\ }\href@noop {} {\bibfield  {journal}
  {\bibinfo  {journal} {Physical Review B}\ }\textbf {\bibinfo {volume} {89}},\
  \bibinfo {pages} {085310} (\bibinfo {year} {2014})}\BibitemShut {NoStop}%
\end{thebibliography}
\providecommand{\noopsort}[1]{}\providecommand{\singleletter}[1]{#1}%

\end{document}